\documentclass[preprint,3p]{elsarticle}
\usepackage[T1]{fontenc}

\usepackage{amssymb}

\usepackage{siunitx}
\usepackage{physics}
\usepackage[hidelinks]{hyperref}
\usepackage{cleveref}
\usepackage{xcolor}
\usepackage{xargs}
\usepackage{soul}
\usepackage{subcaption}
\usepackage{macros}

\usepackage[mode=buildnew,subpreambles=true]{standalone}
\usepackage{shellesc}
\usepackage{tikz,tikzscale}
\usetikzlibrary{arrows,calc,fit,patterns,positioning,spy,backgrounds,decorations.fractals,shapes.misc}
\tikzset{boximg/.style={remember picture,black,thick,draw,inner sep=0pt,outer sep=0pt}}
\usepackage{pgf}
\usepackage{pgfplots}
\usepackage{pgfplotstable}
\usepgfplotslibrary{statistics}
\usepgfplotslibrary{fillbetween}
\usepgfplotslibrary{colormaps}
\pgfplotsset{compat=1.16}
\pgfplotsset{lua backend=true}
\usetikzlibrary{external}
\tikzexternalenable

\usepackage{todonotes}

\begin{document}
\sloppy

\begin{frontmatter}

\title{A framework for high-fidelity particle tracking on massively parallel systems}

\author[ila]{Patrick Kopper\corref{cor}\fnref{fn}}
\ead{kopper@ila.uni-stuttgart.de}
\author[iag]{Anna Schwarz\corref{cor}\fnref{fn}}
\ead{schwarz@iag.uni-stuttgart.de}
\author[boltzplatz]{Stephen M. Copplestone}
\author[iag]{Philip Ortwein}
\author[ila]{Stephan Staudacher}
\author[iag]{Andrea Beck}

\cortext[cor]{Corresponding author}
\fntext[fn]{P. Kopper and A. Schwarz share first authorship.}

\affiliation[ila]{organization={University of Stuttgart, Institute of Aircraft Propulsion Systems},%
            addressline={Pfaffenwaldring 6},
            city={Stuttgart},
            postcode={70569},
            country={Germany}}
\affiliation[iag]{organization={University of Stuttgart, Institute of Aerodynamics and Gas Dynamics},%
            addressline={Pfaffenwaldring 21},
            city={Stuttgart},
            postcode={70569},
            country={Germany}}
\affiliation[boltzplatz]{organization={boltzplatz - numerical plasma dynamics GmbH},%
            addressline={Schelmenwasenstr. 34},
            city={Stuttgart},
            postcode={70567},
            country={Germany}}

\begin{abstract}
Particle-laden flows occur in a wide range of disciplines, from atmospheric flows to renewable energy to turbomachinery. They
generally pose a challenging environment for the numerical prediction of particle-induced phenomena due to their often complex
geometry and highly instationary flow field which covers a wide range of spatial and temporal scales. At the same time, confidence in the
evolution of the particulate phase is crucial for the reliable prediction of non-linear effects such as erosion and fouling. As a
result, the multiscale nature requires the time-accurate integration of the flow field and the dispersed phase, especially in the
presence of transition and separation. In this work, we present the extension of the open-source high-order accurate CFD framework
FLEXI towards particle-laden flows. FLEXI is a massively parallel solver for the compressible Navier-Stokes-Fourier equations
which operates on (un-)structured grids including curved elements and hanging nodes. An efficient particle tracking approach in physical space
based on methods from ray-tracing is employed to handle intersections with curved boundaries.
We describe the models for a one- and two-way coupled dispersed phase and their numerical treatment, where particular emphasis
is placed on discussing the background and motivation leading to specific implementation choices.
Special care is taken to retain the excellent scaling properties of FLEXI on high performance computing infrastructures
during the complete tool chain including high-order accurate post-processing.
Finally, we demonstrate the applicability of the extended framework to large-scale problems.
\end{abstract}


\end{frontmatter}

\section{Introduction}
\label{sec:introduction}

Particle-laden flows with a dilute dispersed phase have long been of scientific interest due to their wide range of occurrences~\cite{Delannay2017,Brandt2022}.
Particles from natural and artificial sources can remain suspended in the surrounding fluid for almost indefinite time and affect a
multitude of disciplines such as pollutant and climate prediction~\cite{Higson1994,Fernando2007,Hefny2009}, human pathogen
transport~\cite{Chang2006,Domino2021}, aeronautical applications~\cite{Ghenaiet2012a,Marx2014,Sommerfeld2021} or
sprays~\cite{McDonald2005,Jones2014}.
Following~\cite{Elghobashi1994}, three distinct classes for the coupling of the fluid and the particle phase can be distinguished, categorizing the coupling into
one-, two-, and four-way coupled approaches.
Both one-way and two-way coupled fluid-particle phases are sufficiently accurate to describe dilute dispersed phases.
The particles are mainly driven by the large scales of the fluid flow, while the reverse influence of the
particles on the flow and thus on the turbulent scales is neglected for one-way coupling, but considered in the two-way
coupled regime.
Following the assumption of a dilute disperse phase, contact forces acting on a particle are only considered through interactions
with solid walls, i.e., inter-particle collisions can be neglected. Dropping the aforementioned assumption leads to four-way coupled phases which resolve collisions but involve a
computationally expensive search for collision partners~\cite{Vance2002}.

As was shown by~\citet{Balachandar2010}, both the Lagrangian point-particle approach and the particle-resolved Direct Numerical
Simulation (DNS) are well suited for the numerical treatment of the considered coupling regimes.
In the Lagrangian point-particle ansatz, an Eulerian field solver for the continuous fluid phase is combined with a scheme for the Lagrangian movement of the
dispersed particles which are modeled as point masses.
The point-particle approximation restricts the application range of the Euler-Lagrange ansatz to particle sizes in the range of the
Kolmogorov scale, i.e., the smallest turbulent flow scale.
A particle-resolved Direct Numerical Simulation (DNS) is recommended for larger particle sizes.
In this approach, the geometry and hence the flow scales around a particle are resolved, thus this ansatz is limited
to small numbers of particles.
Compared to particle-resolved DNS, an Euler-Lagrange (EL) approach is less computationally intensive while still providing individual
particle information which is inevitably lost in Euler-Euler or dusty gas approaches.
However, since the forces induced by the interaction of the particles with the surrounding fluid are not resolved by the Euler-Lagrange
approach as compared to the particle-resolved DNS, they have to be modeled appropriately.
A detailed overview of the numerical treatment of the considered coupling regimes is given in~\cite{Balachandar2010,Kuerten2016}.

Beyond the choice of the coupling regime, a common characteristic of particle-laden flows is the complexity of numerical simulations due to the distinct properties of the different phases and the wide range of encountered scales.
As the accuracy of the particle movement is directly dependent on the resolved scales of the underlying flow field, approaches which aim to reduce the
complexity, e.g., the physical order, of the problem such as the Reynolds-Averaged Navier-Stokes equations (RANS) have shown severe deficiencies in unsteady and turbulent flow fields~\cite{Beck2019}.
Large-Eddy Simulations (LES) offer the required spatial and temporal resolution but come with significant demands towards the computing
resources. Thus, for cases where compromising the solution quality in favor of reduced computational effort is deemed unacceptable,
the focus is placed on the efficient solution of the particle-laden LES on highly parallel systems.
An example of such a case can be found in turbomachinery applications, where the boundary layer dynamics are
unsteady and transition, separation and wake interactions play a dominant role~\cite{Kopper2021a}.

High-order methods are well suited for the accurate solution of such problems due to their inherent low dissipation error.
So far, the EL ansatz has been successfully coupled to high-order finite volume \cite{Kaiser2021}, finite differences
\cite{Patel2022} and discontinuous Galerkin \cite{Beck2019} schemes to enable high-order accurate particle tracking and interpolation in a compressible regime.
In~\cite{Kaiser2021}, the compressible Navier-Stokes-Fourier equations (NSE) are discretized by a $5^\text{th}$-order WENO scheme
which features a local time stepping and a block-based multiresolution ansatz to localize particles efficiently within the Eulerian mesh.
While the framework shows near-optimal parallel performance on highly-parallel systems when restricted to (potentially multiple) continuous phases with a level-set approach~\cite{Hoppe2022},
to the author's best knowledge no scaling results have been published for particle-laden flows.
The authors in \cite{Patel2022} placed less emphasis on efficiency and utilized a finite difference method which solves the volume-filtered NSE with arbitrary order in
space and time.
All frameworks have in common that they are able to capture and handle discontinuities in the solution appropriately.
However, high-order finite volume and finite differences schemes pose challenges to their efficient parallelization due to their inherent non-local solution representation which
results in wide communication stencils stretching across multiple elements. Furthermore, to author's best knowledge both frameworks are not publicly available as open-source.

In order to remedy some of these shortcomings, high-order methods based on a local solution representation have gained significant research interest in recent years, as they offer
excellent scaling properties and are flexible enough to simulate complex geometries.
One such approach is the aforementioned Discontinuous Galerkin (DG) method which was shown to be well-suited for compressible flow simulations involving turbulence and aeroacoustics~\cite{Hindenlang2012,Gassner2012}.
In this work, the open-source high-order accurate CFD framework FLEXI\footnote{\url{https://github.com/flexi-framework/flexi}} is
utilized to solve the compressible Navier-Stokes-Fourier equations using the discontinuous Galerkin Spectral Element Method (DGSEM)
on (un-)structured grids featuring curved faces and hanging nodes.
FLEXI is equipped with pre- and post-processing tools~\cite{Krais2021} and has been successfully applied to a wide variety of
problems including airfoil and turbomachinery simulations~\cite{Beck2018,Kopper2021a} as well as data-driven
shock-capturing~\cite{Beck2020,Zeifang2021}.

In this work, we aim to present a comprehensive description of the extension of FLEXI to incorporate particle tracking
on (un-)structured grids with possibly curved elements.
In addition, since FLEXI is designed to operate efficiently on highly parallel systems, we show that the extended framework preserves
the excellent scaling properties of FLEXI.
The proposed framework features the following characteristics.
Both the approximation of the dispersed and the fluid phase are high-order accurate in space and time for arbitrary orders.
The particle tracking (and the fluid phase) are designed to run efficiently on highly parallel systems and complex geometries
including possibly curved elements and hanging nodes.
The particles can be tracked across discontinuities and shocks in the solution in combination with the shock capturing scheme of
FLEXI~\cite{SonntagPHD,Krais2021}, similar to \cite{Kaiser2021,Patel2022}.
In combination, these features enable an efficient, high-order accurate particle tracking in a compressible carrier phase on arbitrary core counts.
In addition, the presented framework is open-source\footnote{\url{https://github.com/flexi-framework/flexi-particle}}, easily extendable, and can be embedded in machine-learning frameworks
\cite{Kurz2022}.

The focus of this work is on the modeling, challenges and applications of discrete particles embedded in a
continuous flow field. In~\cref{sec:theory}, we present the underlying equations for both the continuous and discrete phase
including particle-wall interactions. We follow by a description of the numerical treatment for these equations in~\cref{sec:methods}.
\Cref{sec:implementation} focuses on the actual implementation, including parallelization, load balancing, and post-processing.
Extensive validation studies are presented in~\cref{sec:validation} before demonstrating the scaling capabilities
in~\cref{sec:parallel}. We show two possible applications of the framework, first the ash deposition  within a low pressure turbine
cascade and second the particle-laden flow around a finite wall-mounted cylinder in~\cref{sec:application}.
We close with a brief conclusion and outlook in~\cref{sec:conclusion}.

\section{Theory}
\label{sec:theory}

\subsection{Continuous Phase}
\label{sec:theory:nse}
The fluid field is governed by the compressible unsteady Navier-Stokes-Fourier equations, given in vectorial form as
\begin{equation}
  \frac{d\cons}{dt} + \nabla\cdot \fphys\left(\cons,\nabla\cons\right) = \source,
\end{equation}
where $\cons = [\rho,\rho u_1,\rho u_2,\rho u_3,\rho e]^T$ is the vector comprising the conservative variables, $\source$ is
a source term, $\rho$ the fluid density, $u_i$ the $i$-th component of the velocity vector and $e$ the total energy per unit mass.
The source term $\source$ accounts for the influence of the dispersed phase on the fluid in two- or four-way coupled regimes.
The physical flux $\fphys$ is composed of the inviscid Euler and the viscous fluxes.
The equation system is closed by the
equation of state of a calorically perfect gas. The dynamic viscosity $\dynvisc$ is obtained from Sutherland's
law~\cite{Sutherland1893}, while the heat flux is given by Fourier's law.
Following Stokes' hypothesis, the bulk viscosity is set to zero.

\subsection{Dispersed Phase}
\label{sec:theory:maxey}
The particles are treated as discrete points which move in a Lagrangian manner according to the following ordinary differential equation
\begin{align}
  \frac{d \partpos}{dt} = \partvel
  \label{eq:part_pos}
\end{align}
with the particle position in physical space $\partpos=[\partpos[][1], \partpos[][2], \partpos[][3]]^T$ and the particle velocity
obtained from the integration of~\cref{eq:theory:maxey}.
The framework is (mainly) intended for simulations which can assume a one-way coupled fluid and dispersed phase, i.e., dilute flows
where the volume fraction is $\volfrac < 10^{-6}$~\cite{Elghobashi1994}.
However, two-way coupling is available under the previous assumption that a particle is a discrete point with zero radius whose influence remains element-local.

The equation of motion for an isolated particle in unsteady uniform fluid is described by the Basset-Boussinesq-Oseen equation,
based on the works by~\citet{basset1888treatise},~\citet{boussinesq1885resistance} and~\citet{oseen1927neuere}.
This equation was deduced under the assumption of unsteady
Stokes flow with a particle diameter $\partdiam$ in the range of the Kolmogorov scale $\nu$ of the surrounding fluid and for small particle
Mach and Reynolds numbers, i.e.,
\begin{align*}
  \rep = \frac{\abs{\fluidvel - \partvel} \partdiam \fluiddens}{\dynvisc} < 1, \
  \map = \frac{\abs{\fluidvel - \partvel}}{\sos} \to 0,
\end{align*}
with the speed of sound $\sos$ of the fluid.
The first attempt towards an equation for unsteady non-uniform fluid flow was derived by~\citet{Tchen1947}. Including the considerations and
improvements of~\citet{Corrsin1956,Auton1988} as well as~\citet{Maxey1983,Gatignol1983}, the resulting
equation is commonly called the Maxey-Riley-Gatignol (MRG) equation. A generalized version which prescribes the particle acceleration in the
Lagrangian frame of reference is given as
\begin{equation}
  \partmass \frac{d \partvel}{dt} = \forcedrag + \fluidmass \frac{D\fluidvel}{Dt} + \frac{\fluidmass}{2}\left(\frac{D\fluidvel}{Dt} -
  \frac{d\partvel}{dt} \right) + \forcebasset + \forcelift + \forcegravity\left(\partmass - \fluidmass\right),
  \label{eq:theory:maxey}
\end{equation}
where $\partmass$ and $\partvel$ are the mass and velocity of the particle, $\fluidmass$ is the mass of the displaced fluid volume,
$\fluidvel$ is the (theoretical) fluid velocity neglecting disturbance caused by the particle itself and interpolated to the particle center of mass, and $\forcegravity$ denotes the gravitational
acceleration.
The substantial derivative of the fluid velocity was introduced for the first time by \cite{Auton1988} and is given as
\begin{align*}
	\frac{D \fluidvel}{Dt}=\frac{\partial \fluidvel}{\partial t}+\fluidvel \cdot \frac{\partial
	\fluidvel}{\partial \mathbf{x}}, \hspace*{1cm}
 \frac{d \fluidvel}{dt}=\frac{\partial \fluidvel}{\partial t}+\partvel \cdot \frac{\partial
	\fluidvel}{\partial \mathbf{x}},
\end{align*}
together with the full derivative along the particle trajectory.
The original MRG equation includes the Fax\'en correction~\cite{Faxen1922a} to account for the local curvature of the
fluid velocity field (non-uniform flow), the influence of which is generally neglected as it is considered to be small compared to the other forces~\cite{Mei1991}.
The terms on the right represent the drag force $\forcedrag$, undisturbed fluid
stresses with contributions from viscous effects and pressure gradients, $\forcepress$, the added mass $\forceamass$, the Basset force $\forcebasset$ encompassing the history term, the combined lift force $\forcelift$ consisting of the Saffman and Magnus force as well as the combined term for
buoyancy and gravity, respectively.
The added or virtual mass considers the work required for the acceleration of the adjacent fluid due to the boundary layer surrounding a particle.
The Basset history term~\cite{Basset1888} causes \cref{eq:theory:maxey} to turn into a fractional-order differential equation, thus inhibiting the use of standard numerical
integration schemes~\cite{Tatom1988,Farazmand2015}.
The approximate integration of this term is prescribed in~\cref{sec:theory:maxey:basset}.
Further modifications have enabled the extension of~\cref{eq:theory:maxey} to higher Reynolds numbers, a detailed overview is
provided in \cite{Crowe2011}, and the compressible regime \cite{Parmar2012}.
As the compressible formulation is restricted to small Mach and Reynolds numbers and for reasons of simplicity, in this work, only the incompressible
form of the MRG equation is considered. Similar to subgrid scale (SGS) models for the continuous phase, the influence of the unresolved scales can be approximated by an appropriate SGS model, e.g.,~\cite{Minier2001,Amiri2006,Breuer2017}.

\paragraph{Stokes Number}
An estimate of the contribution of the right-hand side of~\cref{eq:theory:maxey} on the particle trajectory can be determined with the Stokes number. The Stokes number characterizes the particle behavior in relation to the fluid time scale and is defined as
\begin{align}
  St = \frac{\characvel \partrelaxtime}{\characlength},
  \label{eq:stokesnumber}
\end{align}
which indicates whether a particle is mainly driven by its inertia (higher Stokes numbers) or follows the fluid streamlines (small Stokes numbers).
In~\cref{eq:stokesnumber}, $\partrelaxtime = \smash{\partdiam^2 \partdens (18 \dynvisc)^{-1}}$ denotes the particle relaxation time,
$\characvel$ the characteristic flow velocity and $\characlength$ the characteristic length, see e.g.~\cite{Tropea2007}.

\subsubsection{Drag Force}
\label{sec:theory:maxey:stokes}
The quasi-steady drag force is described exactly through Stokes' law for spherical particles with $\rep < 1$.
To account for higher particle Reynolds numbers, a correction factor $\dragfactor$ is employed, yielding the generalized drag force as
\begin{align}
  \forcedrag = 3\pi\dynvisc \partdiam \dragfactor \left(\fluidvel - \partvel \right)
  \label{eq:theory:maxey:dragforce}
\end{align}
with $f_D \rightarrow 1$ for Stokes flow. The drag factor for spherical particles with higher Reynolds number is obtained from the empirical model of~\citet{schiller1933grundlegenden} as
\begin{align}
  \dragfactor = 1 + 0.15 \rep^{0.687} : \rep < 800.
  \label{eq:dragfactor}
\end{align}
For non-spherical particles, the drag factor is extended to the four-parameter general drag correlation proposed by~\citet{Haider1989} as
\begin{align*}
  f_D = 1 + A \rep^B + \frac{\rep}{24}\frac{C}{1+D/\rep}
\end{align*}
where the parameters $A$ through $D$ depend on the particle sphericity and are obtained from a least squares
fit based on experimental measurements and published in~\cite{Haider1989}.
Following \cite{Loth2008a}, compressibility effects are non-negligible for particle Mach numbers of $\map > 0.6$.
Thus, the drag factor of \cite{Loth2008a}, where $f_D = f(\rep,\map)$, can be employed to account for Mach number influence.

\subsubsection{Saffman Lift Force}
\label{sec:theory:maxey:saffman}
The Saffman lift force accounts for the buoyancy caused by a velocity gradient in the surrounding fluid flow and is modeled after~\cite{Saffman1965,
Saffman1968} as
\begin{align*}
  \forcesaffman = \frac{6.46}{4} C_S \partdiam^2 \sqrt{\frac{\fluiddens \dynvisc}{\abs{\vorticity}}} (\vorticity \times (\urel))
\end{align*}
with the vorticity $\vorticity = \nabla \times \fluidvel$ and the correction of~\cite{Mei1992} for higher Reynolds numbers
\begin{align*}
  C_S =
  \begin{cases}
    1-0.3314\sqrt{\vartheta} e^{-0.1 Re_p} + 0.3314 \sqrt{\vartheta} &: Re_p \leq 40, \\
    0.0524 \sqrt{\vartheta Re_p} &:  Re_p>40,
  \end{cases}
\end{align*}
where $\vartheta = \frac{\partdiam \abs{\vorticity}}{2\abs{\urel}} \ : \ 0.005 < \vartheta < 0.4$.

\subsubsection{Magnus Force}
\label{sec:theory:maxey:magnus}
The Magnus force results from the rotation of particles in motion and is modeled according to~\cite{Rubinow1961} as
\begin{align*}
  \forcemagnus = \frac{\pi}{8} C_M \partdiam^3 \fluiddens (\angularvel \times (\urel)) \frac{\abs{\urel}}{\abs{\angularvel}}
\end{align*}
with an empirical correction by~\cite{Oesterle1998} for higher Reynolds numbers, given as
\begin{align*}
  C_M = 0.45 + \left(4 \frac{Re_\omega}{\rep} - 0.45 \right) \text{exp}(-0.05684 Re_\omega^{0.4} \rep^{0.7}) \ :\\
  \frac{1}{2} < \frac{Re_\omega}{\rep} < 3, \ 10 < \rep < 140.
\end{align*}
Here, the rotational Reynolds number is $\smash{Re_\omega = \frac{\partdiam^2 \abs{\angularvel} \fluiddens}{4 \dynvisc}}$, the relative
fluid-particle angular velocity $\smash{\angularvel=\frac{1}{2}(\nabla \times \fluidvel)  - \vorticity_p}$ and the angular particle velocity $\smash{\vorticity_p=\nabla \times \partvel}$.
The latter results from the temporal integration of an additional ordinary differential equation derived by~\cite{Feuillebois1978}, which
reduces to
\begin{align}
  \moinertia \frac{d \vorticity_p}{dt} = \torque = \fluiddens \frac{\partdiam^5}{64} C_w \angularvel \abs{\angularvel}
  \label{eq:theory:angularvel}
\end{align}
in the steady-state case, where $\torque$ is the torque, $\moinertia = \frac{\pi}{60} \partdens \partdiam^5$ the moment of inertia of a spherical particle and
\begin{align*}
  C_w = \frac{6.45}{\sqrt{Re_w}} + \frac{32.1}{\sqrt{Re_w}} \ : \ 20 < Re_w < 1000
\end{align*}
is a correction factor for higher Reynolds numbers proposed by~\cite{Dennis1980}.

\subsubsection{Basset Force}
\label{sec:theory:maxey:basset}
The history term in the Basset force addresses the temporal delay of the particle boundary layer due to viscous effects, i.e., the acceleration history, and is
given by
\begin{align}
	\textbf{F}_B = \frac{3}{2} \partdiam^2 \sqrt{\pi \fluiddens \mu} \left[\int_{t_0}^t \frac{d (\urel)}{d \tau} (t-\tau)^{-1/2}
    d\tau + \frac{(\urel)\left|_{t_0} \right.}{\sqrt{t-t_0}}\right]
  \label{eq:theory_basset}
\end{align}
with the original Basset kernel~\cite{Basset1888} and a correction term of~\cite{Reeks1983} for particles with
$(\urel)\left|_{t=t_0} \right. \neq 0$.
Following~\cite{VanHinsberg2011}, the derivative in the first term in~\cref{eq:theory_basset} is approximated by a linear term and
the integral is solved by the use of a trapezoidal-based method to handle the singularity in the upper limit.
The number of preceding terms $K$ has to be chosen appropriately to find an trade-off between integration accuracy and memory consumption.

\subsection{Two-Way Coupling}
\label{sec:theory:two_way}
Following the particle-source-in-cell approach proposed by \cite{Crowe1977}, for the two-way coupling, the forces acting on the particles and the corresponding work appear as a source term, $\source =
[0,\sourcei_{m,1},\sourcei_{m,2},\sourcei_{m,3},\sourcei_e]$ in the
momentum equations and the energy equation, respectively.
The source terms for the momentum equations $\source_m=[\sourcei_{m,1},\sourcei_{m,2},\sourcei_{m,3}]$ and the energy equation
$\sourcei_e$ at a point $\vvec{x}_{ijk}, \ i,j,k \in \mathbb{N}_{>0}$ are given by
\begin{align}
  \source_m &= - \mathcal{P}\left\{\left(\forcedrag + \fluidmass \frac{D\fluidvel}{Dt} + \frac{\fluidmass}{2}\left(\frac{D\fluidvel}{Dt} -
  \frac{d\partvel}{dt} \right) + \forcebasset + \forcelift \right), \vvec{x}_{ijk} \right\},\\
  \sourcei_e &= - \mathcal{P}\Big\{\source_m \cdot \partvel , \vvec{x}_{ijk}\Big\},
  \label{eq:theory:source}
\end{align}
with the projection operator $\mathcal{P}\{\cdot,\cdot\}$, which projects the source term onto the grid.
It has to be noted that the assumption of an undisturbed fluid velocity $\fluidvel$ is violated in a two-way coupled ansatz
and recent publications have proposed several approaches to reduce the error introduced through this assumption~\cite{Horwitz2016b}.
Within this work, the influence of the source term is assumed to be restricted to the nearest degree of freedom (DoF), i.e., the projection operator
for an arbitrary variable $a$ is $\mathcal{P}\{a,\vvec{x}_{ijk}\} = \frac{a}{V_{ijk}}$ to ensure conservation, where $V_{ijk}$ is the volume spanned by
the nearest degree of freedom $ijk$.
The particle-induced source term imposes an additional time step restriction, apart from the convective and viscous restrictions.
For its derivation, only the drag force with $\dragfactor=1$ is considered.
The eigenvalue of the resulting ordinary differential equation is $\partrelaxtime^{-1}$, which yields a maximum allowable time step of $dt \leq \partrelaxtime$.
A high-order time integration scheme allows larger time steps, e.g., $dt \leq 2.75
\partrelaxtime$ for a fourth-order Runge-Kutta scheme \cite{Patel2022}.

\subsection{Intersection with Solid Walls}
\label{sec:theory:rebound}
The impact of particles on a wall is approximated by means of a hard sphere model, where the particle intersections are handled in an a posteriori
manner, i.e., $\dtstage > dt_{collision}$, with the time step $dt_{collision}$ required to resolve a collision in
time and the actual time step $\dtstage$.
The change of momentum of a particle between two instances in time is given by
\begin{align*}
  \partmass[2] (\partvel[2]  + 2 (\partvel[1] \cdot \normalvec) \normalvec)- \partmass[1] \partvel[1] = \partmomentum,
\end{align*}
where $(\cdot)_2$ denotes variables after the impact and $(\cdot)_1$ before it.
The normal vector of the boundary is designated by $\normalvec$.
The change of momentum is $\partmomentum = 0$ for a perfectly reflective wall, where a purely elastic deformation of the particle and
the wall is assumed.
A plastic deformation of the particle or the wall results in $\partmomentum \neq 0$.
In this case, the particle properties after impact are determined via coefficients of restitution (CoR); for a quantity $x$ it reads
$\cor{x} = x_2 (x_1)^{-1}$.
CoRs are generally approximated by so-called rebound models based on empirical correlations, often with physical constraints and tunable
parameters which depend on the particles' characteristics and the surface material.
The common rebound models describe the change in the particle trajectory under the assumption that the particle mass
does not change upon impact.
However, the presented framework allows for a change of the particle mass if such a model is available.
Further details on the implemented rebound models are given in~\cite{Tabakoff1981, Bons2017, Whitaker2018, Schwarz2022}.
The change of the angular particle velocity during a wall impact is determined according to~\cite{Crowe2011} as
\begin{align*}
  \moinertia_2 \vorticity_2 - \moinertia_1 \vorticity_1 = - \frac{\partdiam[1]}{2} \left[\normalvec \times (\partmass[2] \partvel[2] - \partmass[1]\partvel[1])\right].
\end{align*}

\section{Numerical Methods}
\label{sec:methods}
In the following, we briefly discuss the numerical treatment of the governing equations for the fluid and dispersed phase.

\subsection{Discontinuous Galerkin Spectral Element Method}
\label{sec:methods:dgsem}
The Navier-Stokes-Fourier equations are solved via the Discontinuous Galerkin Spectral Element Method (DGSEM).
For this, the computational domain $\Omega \subseteq \mathbb{R}^3$ is discretized by non-overlapping, (non-)conforming hexahedral
elements, where the six element faces $\curvedface_k, \ k=1,\ldots,6$ are allowed to be curved.
Curved faces are approximated in a tensor product manner by one-dimensional Lagrange polynomials $l$
up to degree $\ppngeo$ as
\begin{align}
  \curvedface_k(m,n) = \sum_{i,j=0}^{\ppngeo} \curvedface_k(m_i,n_j) l_i(m)l_j(n), \  (m,n) \in [-1,1]^2.
  \label{eq:dgsem_curvedfaces}
\end{align}
Details on the mapping from reference to physical space are provided in~\cite{Hindenlang2015}.

To obtain an efficient discretization scheme, the governing equations are transformed into the reference coordinate system $\boldsymbol{\xi} = [\xi_1,
\xi_2, \xi_3]^T$ of the reference element $E=[-1,1]^3$ via the mapping $\mathbf{x} = \boldsymbol{\chi}(\boldsymbol{\xi},t)$,
$\mathbf{x} \in \Omega$.
The discrete $L_2$ projection onto the test space composed of polynomials $\testfunc(\boldsymbol{\xi})$ up to degree $\ppn$,
followed by an application of Green's identity yields the weak form, given as
\begin{align}
  \int_E J \frac{\partial \cons_h}{\partial t} \testfunc(\boldsymbol{\xi}) d\boldsymbol{\xi} + \int_{\partial E}
  \fnumref \testfunc(\boldsymbol{\xi}) d\vvec{S}
  - \int_E \fphysref(\cons_h, \nabla \cons_h) \cdot \nabla_\xi \testfunc(\boldsymbol{\xi}) d
  \boldsymbol{\xi} = 0,
  \label{eq:dgsem}
\end{align}
with the Jacobian $J$ of the mapping and the outward pointing normal vector $\normalvec$.
In~\cref{eq:dgsem}, $\fphysref$ denotes the contravariant flux vector and $\fnumref$ the numerical flux normal to the element face.
The element-local solution $\cons_h = \cons_h(\boldsymbol{\xi},t)$ is approximated by a tensor product of one-dimensional nodal
Lagrange basis functions $l$ of degree $\ppn$
\begin{align}
  \cons_h(\boldsymbol{\xi},t) = \sum_{i,j,k=0}^{\ppn} \consdofs_{ijk}(t) l_i(\xi^1)l_j(\xi^2)l_k(\xi^3),
  \label{eq:dgsem_interpolation}
\end{align}
with the nodal degrees of freedom $\consdofs_{ijk}(t)$.
\Cref{eq:dgsem} is numerically integrated on the interpolation points by the Legendre-Gauss quadrature with $(\ppn+1)^3$
Legendre-Gauss-Lobatto points.
This collocation of integration and interpolation points allows for a highly efficient scheme.
The Euler fluxes at the cell boundaries are approximated via the numerical flux by Roe, see~\cite{Toro2009}, with the entropy fix by
Harten~\et~\cite{Harten1983b}.
The viscous fluxes are computed with the BR1 scheme of Bassi and Rebay~\cite{Bassi1997}.
The flux is split according to~\cite{Pirozzoli2011, Gassner2016, Flad2017} to mitigate aliasing errors which can lead to stability issues.
Moreover, a suitable shock capturing procedure~\cite{SonntagPHD} is required, as high-order schemes are subject to oscillations in the vicinity of
discontinuities, also known as Gibbs phenomenon.
For this purpose, the DG operator in these cells is replaced by a second-order accurate finite volume (FV) (subcell) scheme with $(\ppn+1)^3$
integral means, which reduces the loss of resolution caused by the inherent higher dissipation of the lower-order FV
operator.
The reader is referred to~\cite{SonntagPHD, Beck2014, Hindenlang2012, Krais2019} for further details on DGSEM, the shock capturing
procedure and applications.
Following the method of lines approach, the explicit low-storage fourth-order accurate Runge-Kutta (RK) scheme by~\citet{Carpenter1994} is employed for the integration in time.
The open-source framework FLEXI\footnote{www.flexi-project.org} is used as a solver for the fluid and dispersed phase which includes the numerical methods mentioned below.

\subsection{Particle Push and Tracking}
\label{sec:methods:tracking}
Particles are advanced in time with the same RK scheme as the fluid phase.
At each stage, the particle push is determined which includes the interpolation of the conserved variables onto the particle
position, the calculation of the corresponding force on the discrete particle, and the subsequent integration of the particle
trajectories in time using the updated particle acceleration. The resulting particle path is obtained through particle tracking
which involves checking for element boundary intersections and subsequent application of appropriate boundary conditions.
The first part of this section will focus on the particle push, while the particle tracking will be discussed in the second one.

\subsubsection{Interpolation}
\label{sec:methods:tracking:interpolation}
The particle position in reference space $\partrefpos = [\xi_p^1,\xi_p^2,\xi_p^3]^T$ enables the interpolation of the conserved
variables onto the center of mass of the particle from which the primitive variables $\prim(\partrefpos,t)=(\rho, u_1, u_2, u_3, p)^T$ can be obtained.
With~\cref{eq:dgsem_interpolation}, the interpolation for a DG element is defined as
\begin{align}
  \cons_h(\partrefpos,t) = \sum_{i,j,k=0}^\mathcal{N} \consdofs_{ijk}(t) l_i(\xi_p^1)l_j(\xi_p^2)l_k(\xi_p^3),
  \label{eq:part_interpolation}
\end{align}
while the second-order FV subcells reduce the particle interpolation to a (linear) interpolation of the
conserved variables to the particle position.
The particle position in reference space $\partrefpos$ is generally determined by an iterative procedure, e.g., via Newton's
method~\cite{Allievi1997}, to find the root of
\begin{align*}
  \boldsymbol{\chi}(\partrefpos) - \partpos = 0.
\end{align*}
This necessitates an initial estimate $\partrefpos^0$ of $\partrefpos$, which can be chosen based on the following three approaches.
The first approach uses the nearest, in the sense of the discrete $L_2$ norm, interpolation point, while the second employs
the nearest mesh point and is available only if curved elements are considered.
Lastly, the third approach is to use the mean distance $\smash{\langle \partpos - \elembary, \sidebary - \elembary \rangle}$,
$i=1,\ldots,6$ between two opposite element faces as initial estimate, where $\elembary$ is the barycenter of the element and $\sidebary$ the barycenter of
the six element faces.
An adequate approach is generally chosen depending on the mesh element type, e.g., the first approach is utilized for linear meshes and the second approach for curved element meshes.

\subsubsection{Integration in Time}
\label{sec:methods:tracking:integration}
In each Runge-Kutta stage, the conserved variables are interpolated onto the particle position using~\cref{eq:part_interpolation}, and
the particle trajectory is obtained by the integration of~\cref{eq:part_pos} and~\cref{eq:theory:maxey} in time.
Hence, the initial particle trajectory (neglecting boundary conditions) in stage $n$, i.e., $t \in [t^n, t^{n+1}]$ describes the path traveled by the particles within $\dtstage=
t^{n+1}-t^{n}$, given by
\begin{align}
  \partpos(t; \alpha) = \partpos(t^n) + \alpha \frac{\parttrajectory}{\abs{\parttrajectory}}, \ \alpha \in
  [0,\abs{\parttrajectory}], \ \parttrajectory = \partpos(t^{n+1}) - \partpos(t^n),
  \label{eq:parttrajectory}
\end{align}
where $\parttrajectory$ denotes the particle trajectory and $\alpha$ the displacement.

\subsubsection{Tracking}
\label{sec:methods:tracking:tracking}
The particle tracking in FLEXI advances particles in time through movement in physical space. While intersection handling is also performed in physical space, the code additionally offers the option to locate particles through an interpolation ansatz in reference space, cf. Ortwein~\et~\cite{Ortwein2019}.

\paragraph{Tracking in Physical Space}
Each element face is checked for an intersection with
$\partpos(t) = \partpos(t^n) + t \partvel$, starting from the element in which the particle resides at $t^n$.
Intersections are considered if $0 < \tintersection \leq \dtstage$, i.e., the particle reaches the face within the current time step.
If the element face is an internal face, the particle is considered to have moved to the adjacent element and the algorithm is
repeated from $\smash{\left.\partpos(t)\right\vert_{t=t^n + \tintersection}}$. If the element face corresponds to a boundary, the corresponding boundary
  condition (open, reflective, periodic) is applied, and the algorithm is continued on the modified trajectory for $dt_\text{remaining} = dt_\text{stage} - \tintersection$.
A special case denotes the intersection with a non-conforming mortar side, where one big element face is matched to two or four small
element faces~\cite{Krais2021}.
Here, only the mapping from small to big faces is unique since the adjacent element can be determined directly.
For the opposite direction, when encountering an intersection with a big mortar side, the code inverts the trajectory and performs the intersection search on the associated small element faces.
Hence, the intersection is again unique and after a second inversion the algorithm can be continued from
$\smash{\left.\partpos(t)\right\vert_{t=t^n + \tintersection}}$ on the small mortar face.

\paragraph{Localization in Reference Space}
This tracking approach inherently does not consider boundary interactions.
In the first step, all particles whose starting positions $\partpos(t^n)$ are near boundaries are again traced in physical space.
Subsequently, the final element for all particles is determined by identifying the particles' host cells with the closest barycenters through a Fast Init Background Mesh (FIBGM, see \cref{sec:implementation:fibgm}) and calculating the position in reference space through Newton's method.
The particle is considered inside the element if $\partpos(\boldsymbol{\xi}) \in [-1,1]^3$, which offers a built-in verification.
The first step is omitted if $\partpos(t^n)$ is far from boundary faces, so that always $\tintersection > dt_\text{stage}$.

\subsubsection{Intersection Handling}
\label{sec:methods:tracking:intersection}
In the following, the treatment of an intersection with an element face is briefly discussed.
Details on the procedure and further literature are given by~\cite{Ortwein2019}.
To efficiently compute an intersection of the particle path with an arbitrary (curved) face, each element face is described by \Bezier
polynomials of degree $\ppngeo$, given in a tensor product manner as
\begin{align*}
  \bezierpoly(\xi,\eta) = \sum_{m=0}^{\ppngeo} \sum_{n=0}^{\ppngeo} \bezierdofs_{mn} \bezierbasis_m(\xi) \bezierbasis_n(\eta)
\end{align*}
with the Bernstein polynomials $\bezierbasis$ of degree $\ppngeo$, the \Bezier control points $\bezierdofs$ and the reference space
$(\xi,\eta) \in [-1,1]^2$.
In order to compute the intersections of the particle trajectory given in~\cref{eq:parttrajectory} with an element face, the roots of
\begin{align}
  \partpos(t>t^n, \displacement) = \partpos(t^n) + \displacement \frac{\vvec{t}}{\abs{\vvec{t}}} \overset{!}{=} \mathbf{p}(\xi,\eta)
  \label{eq:intersection}
\end{align}
have to be found, i.e., $\displacement$, $\xi$ and $\eta$.
In~\cref{eq:intersection}, $\textbf{p}(\xi,\eta)$ denotes the equation of the element face.
However, the evaluation of~\cref{eq:intersection} with $p(\xi,\eta)=\bezierpoly(\xi,\eta)$ is time consuming and, hence, depending
on whether the element faces are curved, more
efficient approaches are applied, as described below.

The type of the element face (planar rectangular, bilinear, planar quadrilateral, curvilinear) is determined by placing a bounding box around their physical extent~\cite{Ortwein2019,Wang2002,Yen1991}.
If this box is empty, the element face is planar, and bilinear or curvilinear otherwise.
The bounding boxes are constructed for each element face at the beginning of the simulation to reduce the computational effort during runtime.
A further benefit is that the bounding box allows to efficiently check whether a particle could possibly intersect an element face or not.
The intersections can then be calculated as follows.

\paragraph{Planar Rectangular Faces}
In the planar case, the particle displacement is calculated according to
\begin{align*}
  \displacement = \frac{(\mathbf{x}_m \cdot \normalvec) - (\partpos(t^n) \cdot \normalvec)}{(\parttrajectory \cdot \normalvec)}
\end{align*}
with the mid point of the \Bezier surface $\mathbf{x}_m$.
The equation of the element face for a planar rectangular face is given by
\begin{align*}
  \mathbf{p}(\xi,\eta) = \mathbf{a} \xi + \mathbf{b} \eta + \mathbf{c},
\end{align*}
where the coefficients $\mathbf{a}$ to $\mathbf{c}$ are based on the corner \Bezier points.
The resulting equation system is solved analytically to obtain $(\xi,\eta)$, and an intersection occurs if $(\xi,\eta) \in
[-1,1]^2$.
\paragraph{Bilinear or Planar Quadrilateral Faces}
The intersection with a bilinear or quadrilateral side is computed according to~\cite{Ramsey2004}, where the surface is described by a bilinear
patch of the form
\begin{align*}
  \mathbf{p}(\xi,\eta) = \mathbf{a} \xi \eta + \mathbf{b} \xi+ \mathbf{c} \eta + \mathbf{d}.
\end{align*}
The parameters $\mathbf{a}$ to $\mathbf{d}$ are determined via the corner \Bezier points of the face, as described
in~\cite{Ramsey2004}, and an intersection is obtained if $\displacement \leq \abs{\parttrajectory}$.

\paragraph{Curved Faces}
To reduce the complexity, the \Bezier surface is projected onto a local coordinate system spanned by two planes which are
orthogonal to each other.
The first plane is defined by the use of the particle trajectory and an arbitrary vector $\mathbf{n}_1$ orthogonal to it.
The second plane is defined in the same way, with the additional condition that the two planes are orthogonal, i.e.,
$\mathbf{n}_2 = \mathbf{n}_1 \times \parttrajectory$.
An intersection with this new 2D plane is handled via \Bezier clipping~\cite{Nishita1990}, a method from ray-tracing, or
Newton's method~\cite{Ortwein2019}.
The latter is applied if the bounding box is flat (curvilinear planar face) and solves $\bezierpoly(\xi,\eta)=0$, i.e., the Jacobian
of the \Bezier polynomials is required.
In \Bezier clipping, the intersections of the particle path with a \Bezier surface, i.e., $\mathbf{p}(\xi,\eta)=\bezierpoly(\xi,\eta)$, are calculated.
For this, two orthonormal vectors are defined by the use of the 2D projected \Bezier control points.
With the help of these vectors, the convex hull of the projected \Bezier points is formed.
This convex hull is utilized to mark regions where intersections are most probable.
The other regions are clipped away with the de Casteljau subdivision, and the procedure is repeated until the intersection point is found.
Further details can be found in the respective literature~\cite{Nishita1990,Ortwein2019}.

\section{Implementation Details}
\label{sec:implementation}
FLEXI is designed to efficiently utilize arbitrary core counts on massively parallel systems. The DG method is well-suited for these
applications since the inter-core exchange of information is limited to the numerical flux through the element faces along the
process boundaries. Additionally, the element-local discretization of DGSEM enables latency hiding through non-blocking
communication. Particles, by contrast, pose additional challenges to efficient parallelization, since both the number of particles
per cell as well as the number of particles crossing the processor boundaries are varying and cannot be determined a priori.
At the same time, immediate communication of particles moving to another processor would be detrimental to performance.
To permit the completion of the particle tracking on the local processor as well as latency-hiding on the discrete phase, FLEXI
follows the halo-region approach to extend the process-local domain.
The section below follows this outline. First, details on the employed parallelization strategy for pure fluid flow are given.
Subsequently, the generation of the halo region and the latency hiding of the runtime communication for both emission and tracking
are presented. After that, a potential load imbalance due to unevenly distributed particles among the processes is addressed.
To preserve the parallel tool chain of FLEXI from pre- to post-processing, the section concludes by presenting the extension of the
post-processing to particles.

It should be noted that the particle tracking is developed in cooperation with the PICLas
framework\footnote{\url{https://github.com/piclas-framework/piclas}} which focuses on solutions to non-equilibrium gas and plasma
flows~\cite{Fasoulas2019,Pfeiffer2018a}.
For an in-depth review on the current state of the particle parallelization approach shared by both codes and their application to
non-equilibrium flows, the reader is referred to~\cite{Kopper2022}.

\subsection{Parallelization}
\label{sec:implementation:parallelization}
The continuous phase in FLEXI is based on the distributed memory paradigm, i.e. each core is assigned and restricted to
its individual memory address space. In order to ensure memory locality and improve cache hit rate, all elements are pre-sorted along a
space-filling Hilbert curve during mesh generation by the open-source mesh generator HOPR~\cite{Hindenlang2015}.
The distribution of the mesh over the cores is performed by utilizing this curve to obtain a continuous segment for each core.
Associated information, such as side connectivity and node coordinates, are
stored non-uniquely along the same curve and are thus also available as contiguous segment.
The space-filling curve approach extends to the
on-disk storage format which allows massively parallel access to non-overlapping data regions through the \textit{HDF5} library~\cite{hdf5}.

Runtime information exchange is handled through non-blocking Message Passing Interface (MPI) communication~\cite{mpi40}. Latency hiding
is extensively used to ensure maximal time intervals available for communication without
stalling the code. Since inter-processor information exchange is limited to direct neighbors which are known a priori,
FLEXI performs in-memory re-ordering of side and node information.
Thus, memory locality is facilitated, and the data is inherently stored in the linear buffers required for MPI communication.

\subsubsection{Halo Region}
\label{sec:implementation:halo}
While the continuous phase is calculated in reference space, particles are tracked in physical space. Subsequently,
the complete geometry information along a path $\left.\mathbf{x}_p\right\vert_{t=t} \rightarrow
  \left.\mathbf{x}_p\right\vert_{t=t+dt_\text{stage}}$
is required to complete their time integration. A halo region provides geometry information within a given distance around the local domain~\cite{Ortwein2019}.
This region permits each core to perform particle tracking until the final particle position is achieved and avoids unnecessary communication during a time increment.

The sorting along a space-filling curve allows for fast domain decomposition, but the position along the SFC provides no information about the cell location in physical space.
As a consequence, an efficient search in physical space must be performed to create the halo region required for a performant particle tracking in the parallel context.
Distributed approaches encounter a severe performance bottleneck on massively parallel systems as each encountered halo element requires grid information local to one processor
to be communicated to a multitude of other processors. To avoid this limitation, the Lagrangian particle implementation is based on
MPI-3 shared memory programming, which results in a hybrid memory code when particles are enabled. For this, the complete raw mesh information,
which is composed exclusively of element information, face connectivity and node coordinates without derived metrics, is stored in a shared memory region on each compute-node.
Based on this information, a communication-free two-step search algorithm is performed to identify halo elements~\cite{Kopper2022}.
These elements are subsequently added to the elements locally considered for particle tracking. While the metrics are restricted to the local and halo
elements, the identifiers from the global mesh are kept in order to ensure consistent numbering throughout the computational domain.
Detailed information on this topic can be found in~\cite{Kopper2022}.

\subsubsection{Cartesian Background Mesh}
\label{sec:implementation:fibgm}
In addition to the halo element search, tracking a particle in physical space necessitates an efficient scheme for the identification of
the element in which the particle resides. As previously stated, the element identification stored for a particular grid element allows
for no correlation to its position in physical space. Thus, the task to correlate the corresponding element to any given position would
involve an elaborate search over potentially the entire grid. To alleviate this problem, a Cartesian Background Mesh (BGM) is created~\cite{Ortwein2019}.
Upon code initialization, the computational domain is overlaid with an I,J,K-identifiable Cartesian grid and the mapping from each BGM cell to all
overlapping unstructured mesh elements is built. This reduces the potential mesh elements associated with each position, ideally down to a single candidate.

\subsubsection{Emission}
\label{sec:implementation:emission}
The information associated with each BGM cell is also utilized in the particle emission step.
During emission, a potentially large number of initial positions in physical space must be mapped
to their corresponding elements. Depending on the quality of the initial guess, this can result in a costly identification step and is thus again parallelized. A processor
takes part in this step if its local mesh region has at least partial overlap with the complete emission region. However, in contrast to tracked particles, there is no
guarantee that an initial position in physical space corresponds to either a local or halo element on a respective processor. Moreover, without having knowledge of the element
ID associated with this particle, a processor is unable to identify the corresponding processor.

To alleviate this problem, the number of associated mesh elements and the affiliated MPI ranks are stored for each BGM cell during the initialization.
Then, the first step after calculating all physical particle positions is to inquire whether the local compute node has all associated mesh elements of
a BGM cell available in shared memory. If this is not the case, the particle positions are collected and sent to all processors associated with the BGM cell
through non-blocking MPI communication. The localization of the remaining particles is then used to hide this communication.
Particles which are localized during this step and unambiguously matched to another processors, i.e. residing either on the same compute-node or in the halo region, are sent exclusively to that core.
In the last step, each processor locates all received particle positions only in the processor-local mesh elements.
Thus, these particles are discarded by all processors except the relevant processor.

\subsection{Latency Hiding}
\label{sec:implementation:latency}
Any communication between processing cores incurs latency costs, which are greatly exacerbated when performing inter-node communication as is generally the case in today's highly parallel systems.
Efficient parallelization thus requires the minimization of the amount of exchanged information while at the same time maximizing
the time available for the completion of the remaining part without stalling the code. DG schemes are well-suited for this task since the volume integral is a
purely local operation which can be utilized to hide the latency for the communication of the cell face information across MPI
borders. A detailed description of the efficient parallelization of the DG operator for the pure fluid phase (including FV subcells) is given in~\cite{Krais2021}.

\begin{figure}[ht]
  \centering
  \includegraphics[width=.8\columnwidth]{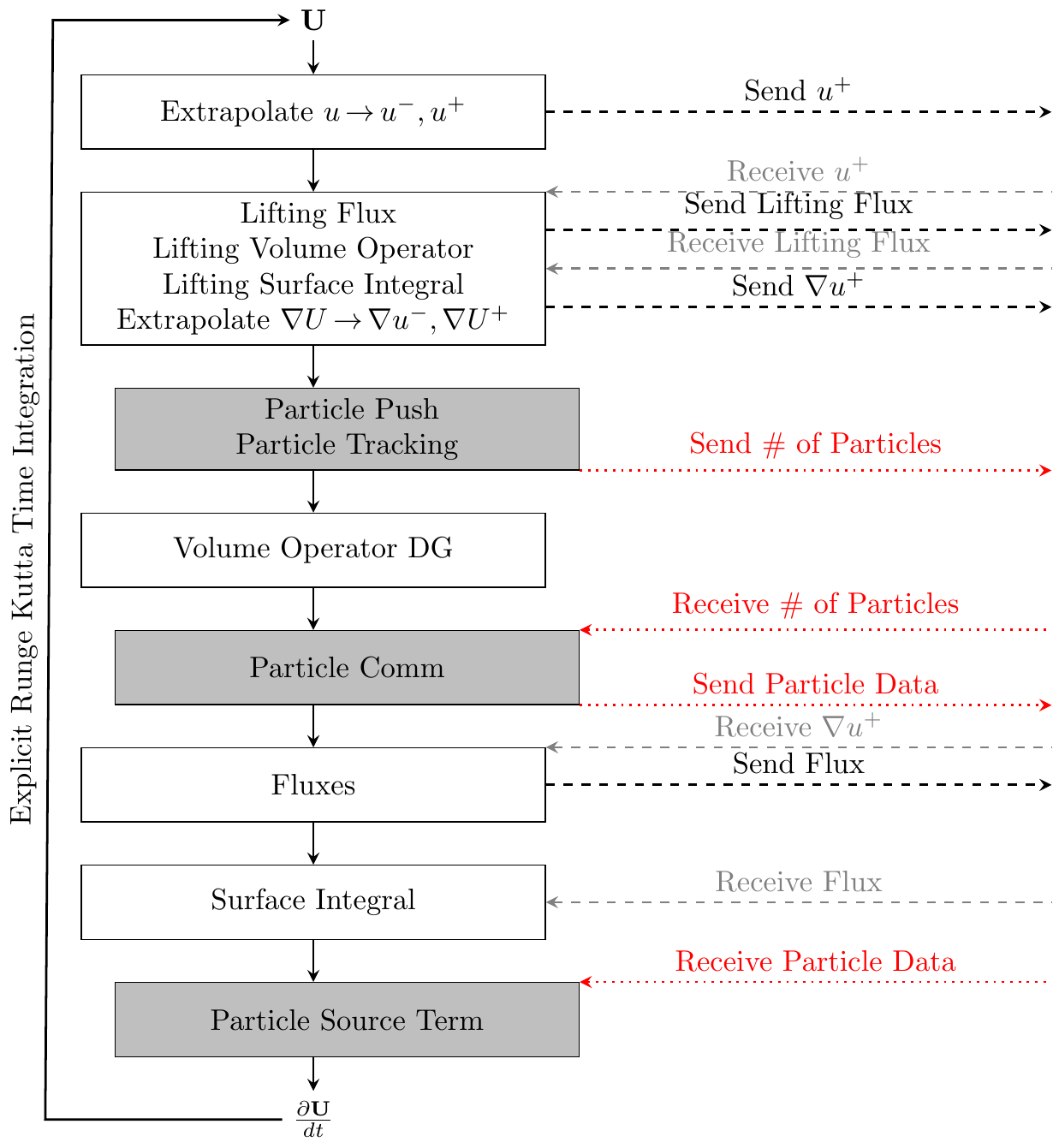}
  \caption{Flow chart of the discontinuous Galerkin operator for particle-laden flow.}
  \label{fig:implementation:latency}
\end{figure}

However, the presence of a dispersed phase adds another challenge towards efficient parallelization as the number of particles
crossing an MPI boundary are generally not known a priori.
This results in the necessity for a two-stage communication, as depicted in~\cref{fig:implementation:latency}. In the first step, only the number of exchange
particles is communicated, thus enabling the receiving core to open the corresponding MPI buffers. Clearly, this step is latency-dominated rather
than bandwidth-dominated. For this reason, we elected to hide this step behind the volume integral that is also utilized for the latency hiding of the fluid
phase. The second step involves the transfer of the actual particle information. Since this step is more bandwidth-intense, the communication is hidden
behind the surface integral, which avoids a stacking on the DG communication and maximizes the time before data starvation occurs.
The result is a highly-efficient operator for particle-laden flow.

\subsection{Load Balance}
\label{sec:implementation:loadbalance}
Pure DG schemes generally require a fixed effort per grid element. While the exact computing time will be machine dependent, it is
sensible to presume that it remains constant across the machine.
Given the additional assumption of negligible stalling through communication latency, an ideal load balance distribution can be straightforwardly achieved by evenly distributing the elements among the processors.

However, the distribution of the Lagrangian particles is generally not known a priori. Additionally, the ratio of computational time
required to advance one degree of freedom in time compared to a particle integration step is machine-dependent. Thus, an efficient
load balancing must take both the varying particle distribution and the ratio of computing efforts into account. FLEXI follows a
warm restart-based load balancing approach~\cite{Ortwein2018}. Runtime measurements are performed at fixed time intervals to determine
the actual CPU time, which is then assigned to the individual mesh elements. If a sufficiently high imbalance is detected, load
balancing is performed through in-memory redistribution of the elements along the space-filling curve. While this procedure does not
seek to minimize communication as, e.g., graph-based distributions attempt, the domain decomposition step is generally faster
compared to more advanced approaches~\cite{Schamberger2003,Mitchell2007}, and the resulting distribution performs similarly given sufficient latency hiding.

\subsection{Post-Processing}
\label{sec:implementation:postprocessing}
In order to visualize our results in a highly parallel manner, we extended the custom-built visualization tool chain described
in~\cite{Krais2019} with respect to particles.
This tool chain can be used in combination with the open-source software ParaView~\cite{ahrens2005paraview}.
The interested reader is referred to~\cite{Krais2019} for details on the post-processing and visualization tools for the fluid phase.
Our post-processing tools include a plugin for ParaView written in C/C+ and a standalone visualization tool written in Fortran.
In addition to the fluid flow, both methods overlay the particle location in the domain and any historical information of any particle that passed a
boundary condition or impacted upon a wall since the last write-out.
In order to keep this information synchronized, the instantaneous particles' characteristics as well as historical impact data are written at the same time and to the identical \textit{HDF5} file as the nodal fluid field information.
For the instantaneous particles, the particle position in physical space, the particle velocities, the species and the number of
impacts with a wall are written.
The old and the new particle trajectories, velocities and kinetic energies as well as the species, the impact time, the boundary it
crossed and the number of reflections so far are saved for the impacting particles while we chose to visualize these particles on their
impact position.
Additional particle information can be saved if provided by their respective models outlined in~\cref{sec:theory:maxey}.

\section{Validation}
\label{sec:validation}
Before turning to actual application cases, we validate the various building blocks for particle-laden flow, from temporal integration to two-way coupling.
For the validation of the fluid phase, see e.g.~\cite{Beck2014,Beck2016,Beck2018}.
In the following, all forces in~\cref{eq:theory:maxey}, except the drag force, are neglected, the drag factor in
\cref{eq:dragfactor} is utilized and one-way coupling is assumed, unless stated otherwise.

\subsection{High-Order Time Integration}
\label{sec:validation:testcases}
First, the high-order time integration of the particles is validated.
For this, the particle transport and momentum equations given
in~\cref{eq:theory:maxey} and~\cref{eq:part_pos} were analytically integrated in time for $\dragfactor=1$.
A stationary flow field with a linear velocity profile was considered, i.e., $\rho=p=1$ and $\vel=[y/\partpos[][2]|_{t=0},0,0]^T$.
The particles were initialized at $\partpos=[0,2,0.5]^T$ with $\partvel = \textbf{0}$, and the particle momentum was $\partmass \frac{d \partvel}{dt} =
\forcedrag [1,0,0]^T + \partmass \forcegravity$ with $\forcegravity=[0,g,0] = [0,-9.81,0]^T$.
The fluid velocity at the particle position $\fluidvel=[\partpos[][2]/\partpos[][2]|_{t=0},0,0]^T$ is based on the particle path in $y$-direction due to the linear velocity
profile and normalized by $\partpos[][2]|_{t=0}=2$ which yields $\fluidveli{1} = \frac{\partpos[][2]}{\partpos[][2]|_{t=0}} =
1+\frac{g t^2}{2\partpos[][2]|_{t=0}}$.
Thus, the analytical integration results in
\begin{align*}
  \partvel[][1,\text{ex}] =& \ 1-e^{-\frac{t}{\partrelaxtime}} + \frac{g}{2 \partposi{2}|_{t=0}} \left(t^2 - 2 t \partrelaxtime + 2\partrelaxtime^2 -
    2\partrelaxtime^2 e^{-\frac{t}{\partrelaxtime}}\right),\\
  \partpos[][1,\text{ex}] =& \ t - \partrelaxtime\left(1-e^{-\frac{t}{\partrelaxtime}}\right) + \frac{g}{2 \partposi{2}
  |_{t=0}}
  \left(t^3 - t^2 \partrelaxtime + 2\partrelaxtime^2t +
  2\partrelaxtime^3 \left(e^{-\frac{t}{\partrelaxtime}} - 1 \right)\right),\\
  \partvel[][2,\text{ex}] =& g t + \partvel[][2]|_{t=0}, \
  \partpos[][2,\text{ex}] = \frac{1}{2} g t^2 + \partpos[][2]|_{t=0}.
\end{align*}
The study was preformed on a computational domain of $\Omega = [0,2]^3$ discretized with two elements in each direction.
Three particles were investigated, each with a different Stokes number $St = \{0.1,1,10\}$, until $t=0.4$ with $dt = \{0.0125 \cdot 2^k: k \in
\mathbb{N}, k \in [0,4]\}$.
The particle relaxation time was calculated via~\cref{eq:stokesnumber}, $\characvel=y/2$ and $\characlength=2$.
\Cref{fig:validation:timeintegration} depicts the discrete $L_1$ norm between the analytically, $(\cdot)_\text{ex}$, and
numerically, $(\cdot)_\text{num}$, integrated particle
position, defined as $L_1 = \abs{\partpos[p,\text{ex}] - \partpos[p,\text{num}]}$ and the theoretical slope.
The $4^\text{th}$-order accuracy in time is achieved for all Stokes numbers considered in this study under the condition that $\dragfactor=1$.

\subsection{Particle Push}
\label{sec:validation:RHS}
\begin{figure}[htpb]
  \begin{subfigure}[t]{.51\textwidth}
    \begin{flushleft}
      \includegraphics[width=.8\columnwidth]{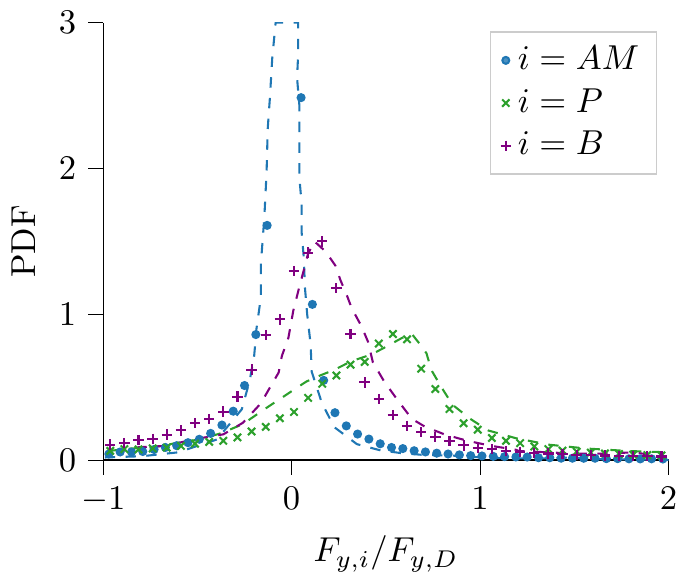}\vspace*{-\baselineskip}
    \end{flushleft}
    \caption{$\partdiam = 0.005 \delta$}
  \end{subfigure}%
  \hfill
  \begin{subfigure}[t]{.475\textwidth}
    \begin{flushleft}
      \includegraphics[width=.8\columnwidth]{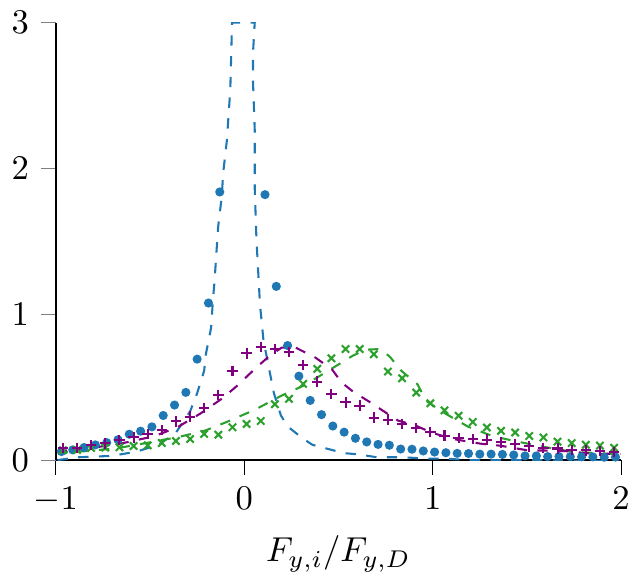}\vspace*{-\baselineskip}
    \end{flushleft}
    \caption{$\partdiam = 0.01 \delta$}
  \end{subfigure}
  \caption{Discrete Probability Density Function (PDF) of the forces $F_{y,i}$ acting on the particles in wall-normal direction. The forces are normalized by the
  drag force in wall-normal direction. The dashed lines highlight the results of~\cite{Armenio2001}.}
  \label{fig:validation:RHS}
\end{figure}

Following~\cite{Armenio2001}, the effect of the forces, $\forcedrag$, $\forcepress$, $\forceamass$ and $\forcebasset$, on a
particle is quantified by the use of a turbulent channel flow at $Re_\tau =
175$ and $Re=4050$ laden with particles of two different Stokes numbers.
The particle density is set to $2.65 \rho$, $\rho=1$, and particle diameters of $\partdiam = 0.005 \delta$ and $\partdiam = 0.01 \delta$
were investigated, where $\delta=1$ is the channel half height.
For each particle diameter, three species were initialized, one for each of the forces, $\forcepress$, $\forceamass$ and
$\forcebasset$.
The computational domain $\Omega = [0,4 \pi \delta] \times [0,2 \delta] \times [0,4 \pi \delta /3]$ was discretized by $x$ elements
with $\ppn=5$.
A total of \num{3072} particles were uniformly emitted at four planes distributed equidistantly along the streamwise direction.
Since a fixed time step $dt$ is required for the temporal integration of the Basset force~\cite{VanHinsberg2011}, the particles were advanced in
time by the explicit Euler scheme with $dt=\eh{-5}$.
As the authors in~\cite{Armenio2001} omit the description of the numerical treatment of the Basset force, the number of previous time steps
was chosen as $K=20$ and $K=100$ for $\partdiam = 0.005 \delta$ and $\partdiam = 0.01 \delta$, respectively.
Particles were emitted at $4 T^*$ and statistics have been accumulated over $0.1 T^*$, with the characteristic time
$T^*=\frac{\delta}{u_{\tau}}$.
As depicted in~\cref{fig:validation:RHS}, the results are in agreement with the literature.

\subsection{Cylinder in Cross-Flow}
\label{sec:validation:cylinder}
\begin{figure}[htpb]
  \begin{subfigure}[t]{.49\textwidth}
    \begin{flushleft}
      \includegraphics[width=.8\columnwidth]{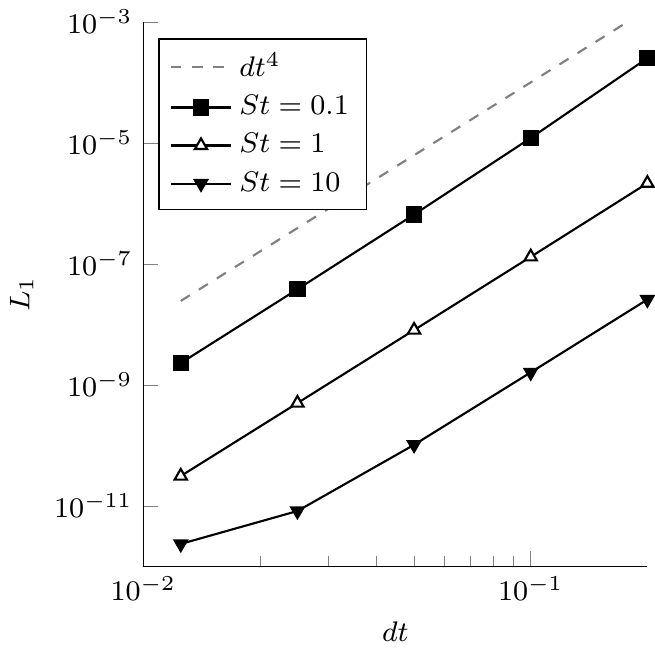}\vspace*{-\baselineskip}
    \end{flushleft}
    \caption{Discrete $L_1$ error between the analytically and numerically integrated particle position in $x$-direction.}
    \label{fig:validation:timeintegration}
  \end{subfigure}%
  \hfill
  \begin{subfigure}[t]{.49\textwidth}
    \begin{flushleft}
      \tikzsetnextfilename{fig_cylinder_impaction}
      \includegraphics[width=.8\columnwidth]{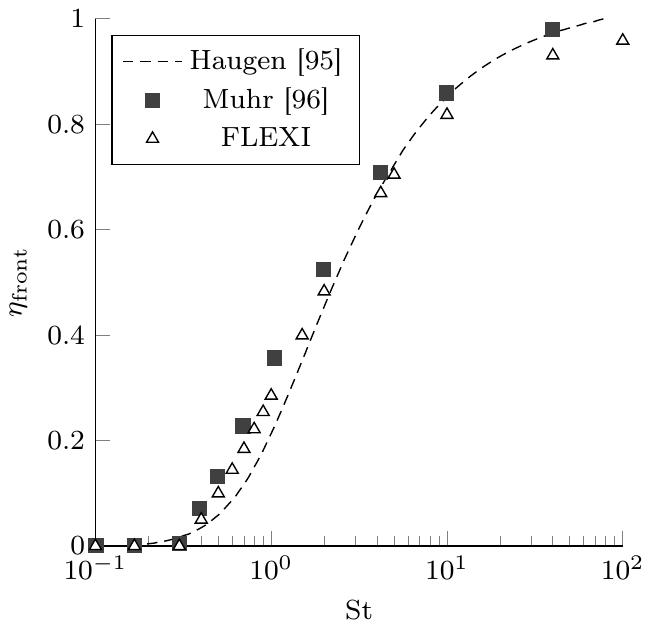}\vspace*{-\baselineskip}
    \end{flushleft}
  \caption{Impaction efficiency over Stokes number for the $Re_D = 491$ cylinder.}
  \label{fig:validation:cylinder:impaction}
  \end{subfigure}
  \caption{Validation for the discrete phase.}
  \label{fig:validation}
\end{figure}

Haugen and Kragset\,\cite{Haugen2010} have shown that the impaction efficiency of a cylinder in a particle-laden crossflow is well-suited for code validation.
The impaction efficiency denotes the ratio of emitted particles which travel in the direction of the cylinder to the number of particles that actually collide with it.
They identified particles with small Stokes numbers to be most sensitive since those show the strongest influence on changes in
the boundary layer while performing their simulations with Reynolds numbers $Re_D$ ranging from \numrange{20}{6600} based on the
cylinder diameter $D$. This validation study follows their example by focusing on the $Re_D = 421$ case.
\Cref{fig:validation:cylinder:impaction} compares the results generated by FLEXI with the data from Haugen and Kragset as well as theoretical considerations of a cylinder at $Re_D = 491$ by Muhr\,\cite{Muhr1976}. FLEXI shows excellent agreement with the literature with a slight discrepancy for high Stokes numbers.

\subsection{Two-Way Coupling}
\label{sec:validation:2wcoupling}
\begin{figure}[htpb]
  \includegraphics[width=.5\columnwidth]{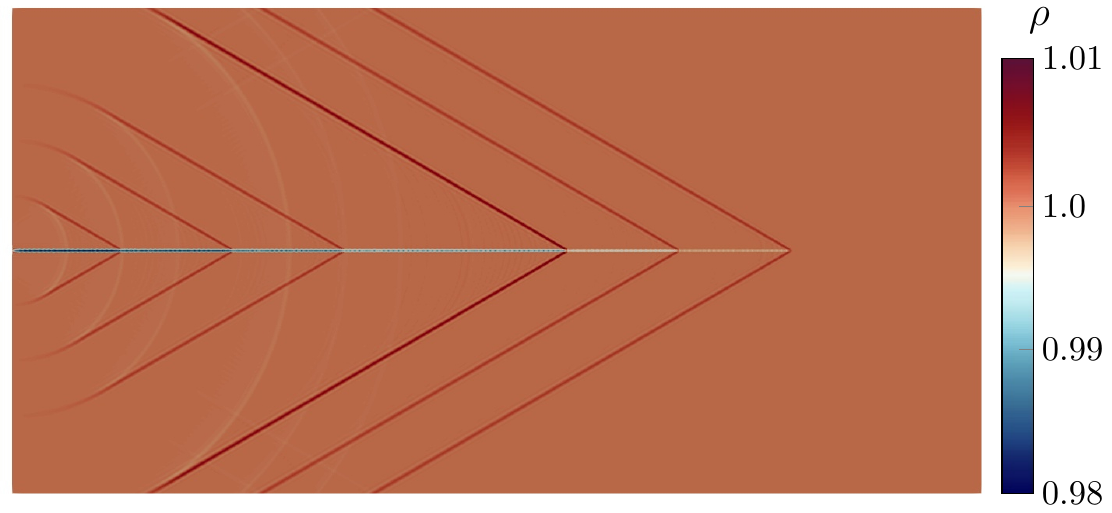}
  \includegraphics[width=.5\columnwidth]{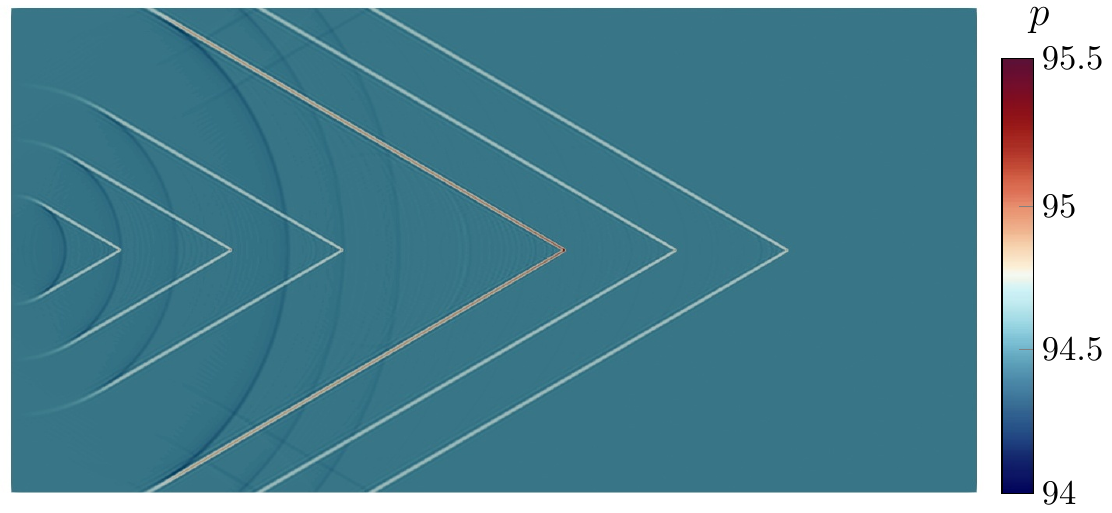}
  \caption{Validation of the two-way coupling. Instantaneous density (left) and pressure profile (right) at $t=0.7$.
  \label{fig:validation:two_way}}
\end{figure}

A comparison of the Mach angle of a supersonic particle traveling through a quiescent fluid with the theoretical value is a suitable test
case to validate the source term introduced by two-way coupled Euler-Lagrangian particles.
For this, a computational domain of $\Omega = [0,2] \times [0,1] \times [0,1]$ was discretized by $199 \times 99 \times 1$ elements with
$\ppn=5$.
The fluid was initially at rest, i.e., $\rho = 1$, $p = 94.464286$, $\vel=\mathbf{0}$ and $\dynvisc=\eh{-5}$, and the particles
moved continuously into the domain at $\partpos = [0.0,0.5,0.5]$ with
$\partvel = [23,0,0]$, resulting in a particle Mach number of $M_p=\frac{\abs{\fluidvel-\partvel}}{c}=2$ with the speed of sound
$c=\sqrt{\kappa \frac{p}{\rho}}$.
The Mach angle $\theta$ is given by $\sin(\theta) = \frac{1}{M_p}$, resulting in $\theta = 30^\circ$, which is reproduced by the
numerical results, as depicted in \cref{fig:validation:two_way}.

\section{Parallel Performance}
\label{sec:parallel}
Following the validation of the framework, we now demonstrate the scaling capabilities of the extended FLEXI and the effectiveness of load balancing.

\subsection{Scaling}
Since FLEXI is designed as massively parallel code, scaling must be retained throughout the framework.
In this work, the scaling was tested on a Cartesian box with elementary dimensions $2L\times L \times L, \ L \in \mathbb{R}_{>0},$ with a freestream flow field and approx.~\num{175000} particles per unit cube.
The domain size was then extended with increasing processor counts to maintain an equal workload per core.
To increase the practical applicability of the test case, additional code features were enabled. The flow and particles were initially convected with
mean velocity $\fluidvel = \partvel = [1,1,1]$ while synthetic turbulence was generated in the first half of the domain through a
Recycling Rescaling Anisotropic Linear Forcing (RRALF) method~\cite{Kuhn2020}. A sponge zone was applied in the final decile to dampen
fluctuations before reaching the outflow boundary. The polynomial degree was set to $\ppn=\num{5}$, resulting in a sixth-order
accurate scheme. Simulations were performed on the HPE Apollo \textit{Hawk} system at the High Performance Computing Center (HLRS) in Stuttgart with dual-socket AMD EPYC\textsuperscript{TM} nodes (\num{128} cores per node) and an InfiniBand HDR200 interconnect. The code was compiled with the GNU compiler version 9.2.0 with the libraries mpt 2.23, hdf5 1.10.5 and aocl 3.0. Each run was repeated \num{3} times to eliminate fluctuations in overall machine load.

\begin{figure}[htpb]
  \begin{subfigure}[t]{.49\textwidth}\vskip 0pt
    \includegraphics[width=.8\columnwidth]{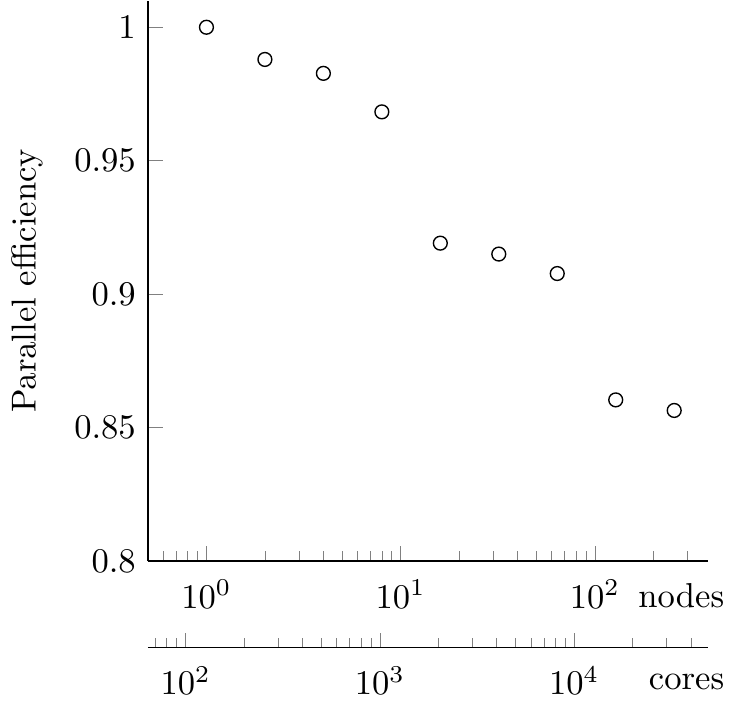}%
    \caption{Weak scaling with \num{6912} DoFs/core and \num{1562} particles/core.}
    \label{fig:scaling:weak}
  \end{subfigure}\hfill%
  \begin{subfigure}[t]{.49\textwidth}\vskip 0pt
    \includegraphics[width=.8\columnwidth]{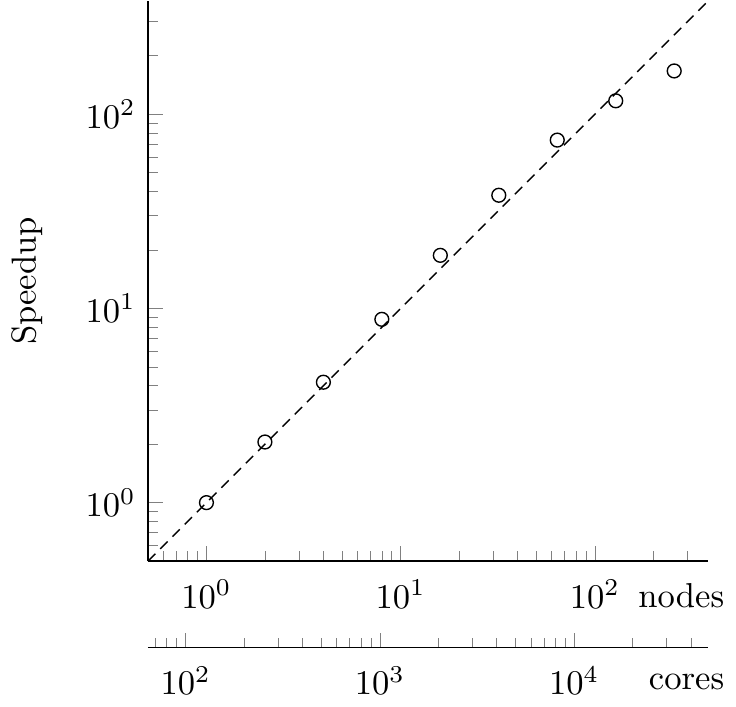}%
    \caption{Strong scaling with \num[exponent-product = \cdot]{5.66E7} DoFs and \num[exponent-product = \cdot]{1.28E7} particles.}
    \label{fig:scaling:strong}
  \end{subfigure}
  \caption{Parallel efficiency for the structured rectangular mesh.}
  \label{fig:scaling}
\end{figure}

Parallel efficiency in terms of weak and strong scaling is depicted in~\cref{fig:scaling}. The interconnect on \textit{Hawk} is deployed in a 9-dimensional enhanced hypercube topology resulting in diminishing bandwidth as the node number increases. This is observable as jump in parallel efficiency with the performance leveling out again for higher node counts. Nonetheless, FLEXI demonstrates excellent scaling with parallel efficiency for the weak scaling case at $>\SI{85}{\%}$ including for the \num{256} nodes (\num{32768} core) run. Higher core numbers are possible but require changing some internal counters to \num{8} byte-integers which was omitted for this tests. Strong scaling temporarily even exceeds the ideal speedup due to relieved memory pressure and the additional latency hiding capability introduced with the particle load.

\subsection{Load Balancing}
\begin{figure}[htpb]
  \begin{subfigure}[t]{.49\textwidth}
  \begin{center}
    \includegraphics[width=\columnwidth]{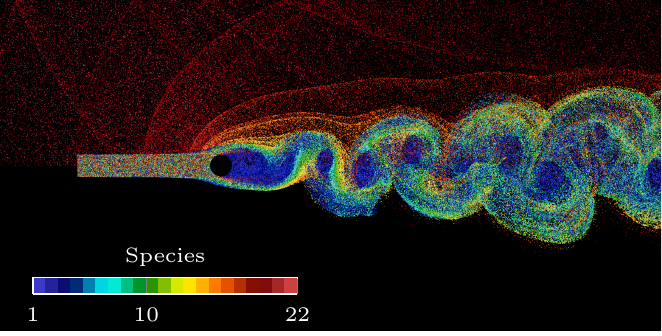}
    \caption{Instantaneous particle positions, top: with reflected particles, bottom: without reflected particles.}
    \label{fig:validation:cylinder:particles}
  \end{center}
  \end{subfigure}\hfill%
  \begin{subfigure}[t]{.49\textwidth}
  \begin{center}
    \includegraphics[width=\columnwidth]{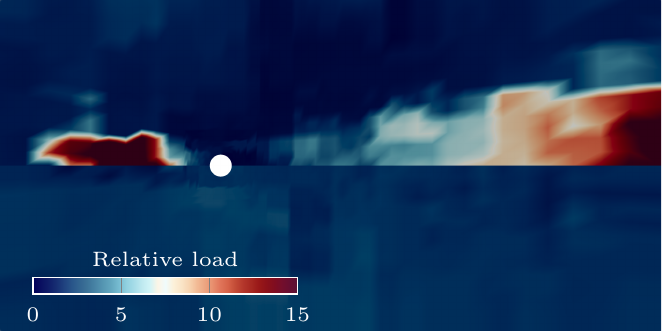}
    \caption{Average load per processor, top: before load balancing, bottom: after load balancing.}
    \label{fig:validation:cylinder:loadbalance}
  \end{center}
  \end{subfigure}
  \caption{Instantaneous particle positions and load distribution for the $Re_D = 491$ cylinder.}
  \label{fig:validation:cylinder}
\end{figure}

Load balancing is evaluated using the setup from~\cref{sec:validation:cylinder}. Instantaneous particle positions colored by species
are shown in~\cref{fig:validation:cylinder:particles}. While only the non-reflected particles are necessary for the validation
in~\cref{sec:validation:cylinder}, we opted to keep reflected particles in order to further validate the boundary intersection
accuracy. \Cref{fig:validation:cylinder:loadbalance} depicts the average load per processor, sampled across \num{100} iterations,
with red areas denoting high relative computational load and blue areas indicating minimal loading. In the top half, the load before load balancing
is illustrated, where most of the computational effort is focused on the axial direction upstream and downstream of the cylinder position. The load drops in the vicinity of the cylinder as the grid element size decreases towards the cylinder. The bottom half depicts the load after balancing using the same scale.
While the distribution is not perfectly uniform, it has moved considerably closer to equilibrium.

\section{Applications}
\label{sec:application}
In this section, we briefly demonstrate the applicability of FLEXI to more challenging large-scale test cases, the ash deposition in a turbine linear cascade
and the particle-laden flow around a wall-mounted cylinder.
In the following, all forces except the drag force are neglected and the drag factor in~\cref{eq:dragfactor} is employed, unless stated otherwise.
Furthermore, a one-way coupled fluid and dispersed phase is assumed.

\subsection{Turbine Linear Cascade}
\label{sec:application:t106c}
\begin{figure}[htpb]
  \begin{subfigure}[b]{.5\textwidth}
    \begin{subfigure}[b]{\textwidth}
      \begin{flushleft}
        \includegraphics[width=\columnwidth]{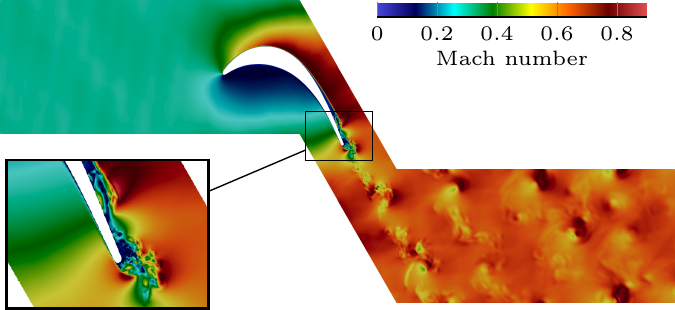}\vspace*{-\baselineskip}
      \end{flushleft}
      \caption{Mach number distribution.}
      \label{fig:application:t106c:mach}
    \end{subfigure}\\
    \begin{subfigure}[b]{\textwidth}
      \includegraphics[width=\textwidth]{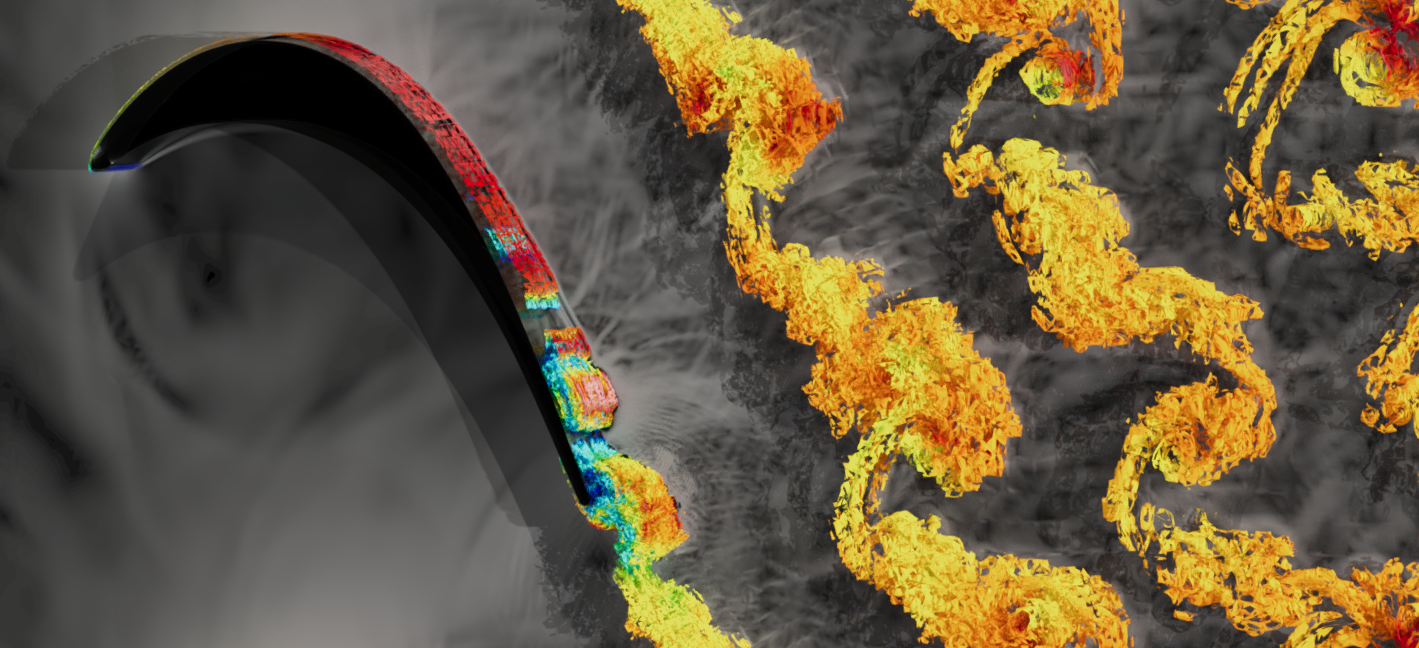}
      \caption{Instantaneous isosurfaces of the Q-criterion colored by Mach number with Schlieren visualization in the background.}
    \end{subfigure}
  \end{subfigure}\hfill%
  \begin{subfigure}[b]{.5\textwidth}
    \begin{flushright}
      \includegraphics[width=.98\columnwidth]{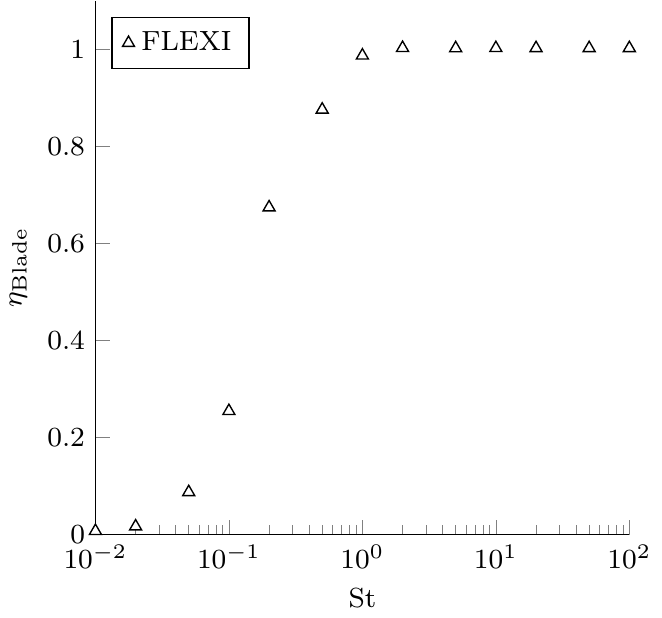}\vspace*{-\baselineskip}
    \end{flushright}
  \caption{Impaction efficiency over Stokes number.}
  \label{fig:application:t106c:impaction}
  \end{subfigure}
  \caption{T106C low-pressure turbine cascade results.}
  \label{fig:application:t106c}
\end{figure}

Ash deposition was identified as a major contribution to performance degradation of turbines, especially on modern specimens with hot
section temperatures of or exceeding \SI{1350}{\kelvin}~\cite{Dunn1996}. In this application, we simulate the ash deposition on a
T106C low-pressure turbine linear cascade.
The setup was chosen as described by~\cite{Hillewaert2013} with an exit Mach number of $M = 0.65$ and Reynolds number of
$Re = \num{80000}$. The mesh is fully periodic in pitchwise and spanwise direction with the Mach number distribution shown
in~\cref{fig:application:t106c:mach}. More details on the mesh and fluid setup can be found in~\cite{Beck2019}. Ash density was
estimated at $\partdens=\SI{990}{\kilogram/\metre^3}$, and \num{16} particle species with Stokes numbers ranging from \numrange{0.01}{1000.} were injected over the entire inlet
boundary with the local fluid velocity. As the deposition rate is highly variable and thus strongly dependent on the chosen model,
only the initial impaction efficiency is illustrated in~\cref{fig:application:t106c:impaction}. Note that similar
to~\cref{fig:validation:cylinder:impaction}, the impaction efficiency will approach unity as the shading region of the blade extends
across the complete passage for heavy particles.

\subsection{Wall-Mounted Cylinder}
\label{sec:application:cylinder}
Particles suspended in the ingested air are a major source of fan and compressor erosion of jet engines.
This more challenging setup investigates the particle-laden flow around a wall-mounted cylinder at $Re_D = \num{32000}$ which is representative of
the leading edge of a transonic compressor blade. The numerical setup is based on experiments by Kawamura~\cite{Kawamura1984} who
evaluated wall-mounted cylinders at varying height, $H$, to diameter, $D$, ratios $H/D$. For this study, the domain of the $H/D = 8$ case
was discretized using a fully hexahedral mesh with \num{788112} elements and $\ppn = 7$.
In order to retain the high-order geometry near the cylinder surface, the mesh features full volume curving with $\ppngeo = 4$ using agglomeration to generate high-order inner element mappings.
The turbulent inflow boundary layer was generated using the RRALF approach.

\begin{figure}[htbp]
  \centering
  \includegraphics[width=\textwidth]{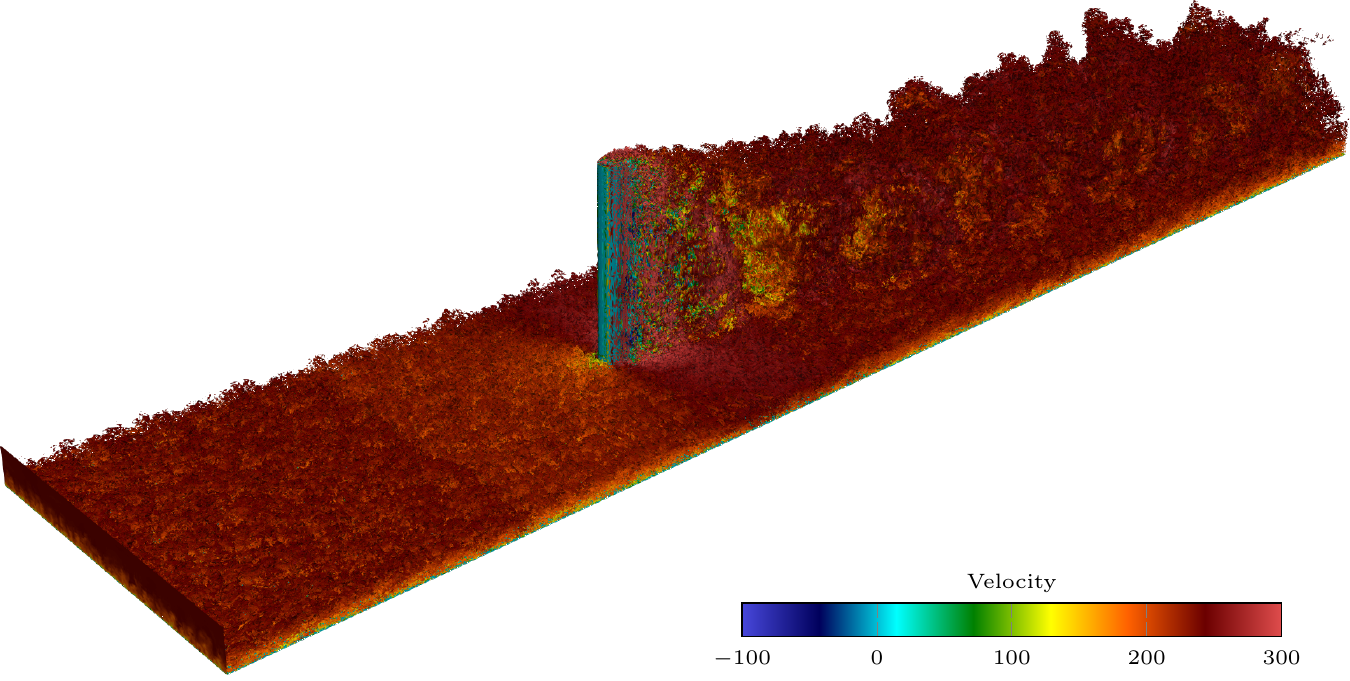}
  \caption{Instantaneous isosurface of the Q-criterion colored by downstream velocity.}
  \label{fig:application:cylinder:qcrit}
\end{figure}

The generated inflow turbulence with an inlet Mach number of $M = \num{0.7}$ is illustrated in~\cref{fig:application:cylinder:qcrit} using the instantaneous isosurface of the Q-criterion colored by
the velocity component in downstream direction.
The RRALF region covers the left side of the graph, distinguishable by a small discontinuity in the velocity magnitude due to the upstream pressure field of the cylinder.
The RRALF is shown to produce a statistically stable boundary layer flow with a Reynolds number of $Re_\theta = \num{3299.12}$ based on momentum thickness at the cylinder position.

\begin{figure}[htbp]
  \begin{subfigure}[b]{.29\textwidth}
  \includegraphics[width=\textwidth]{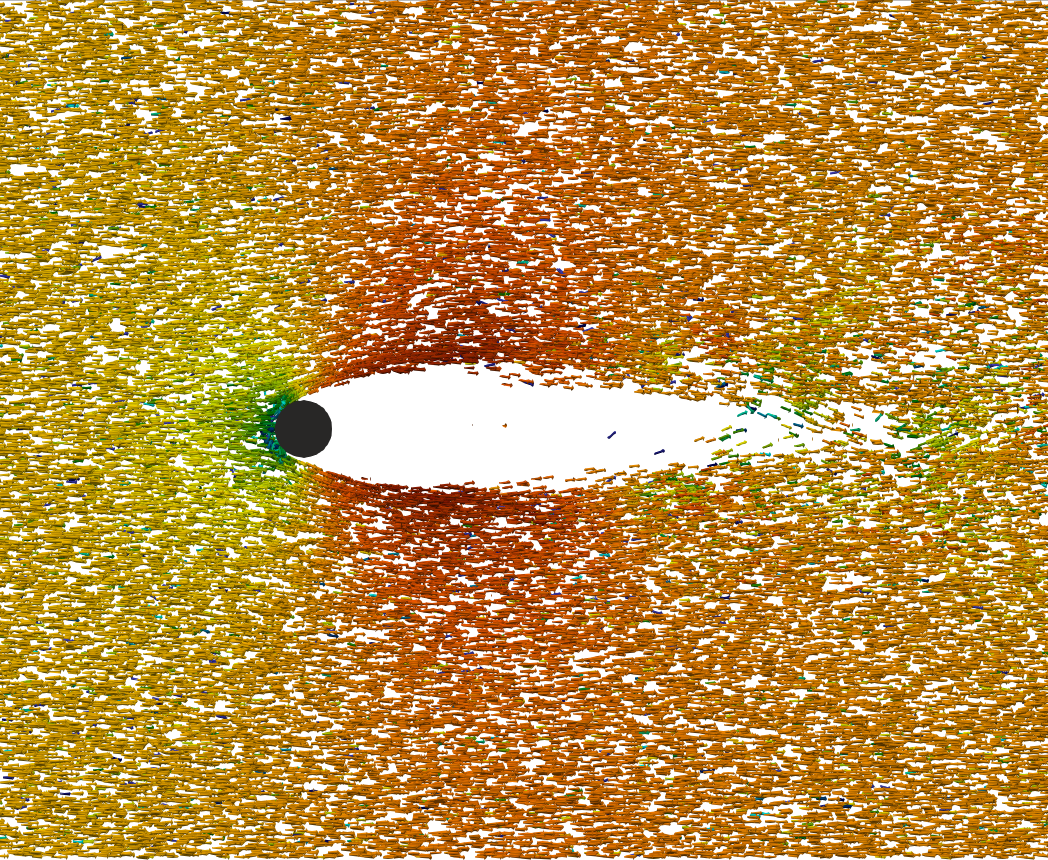}
  \caption{$St = \num{0.1}$}
  \label{fig:application:cylinder:dist:st01}
  \end{subfigure}\hfill
  \begin{subfigure}[b]{.29\textwidth}
  \includegraphics[width=\textwidth]{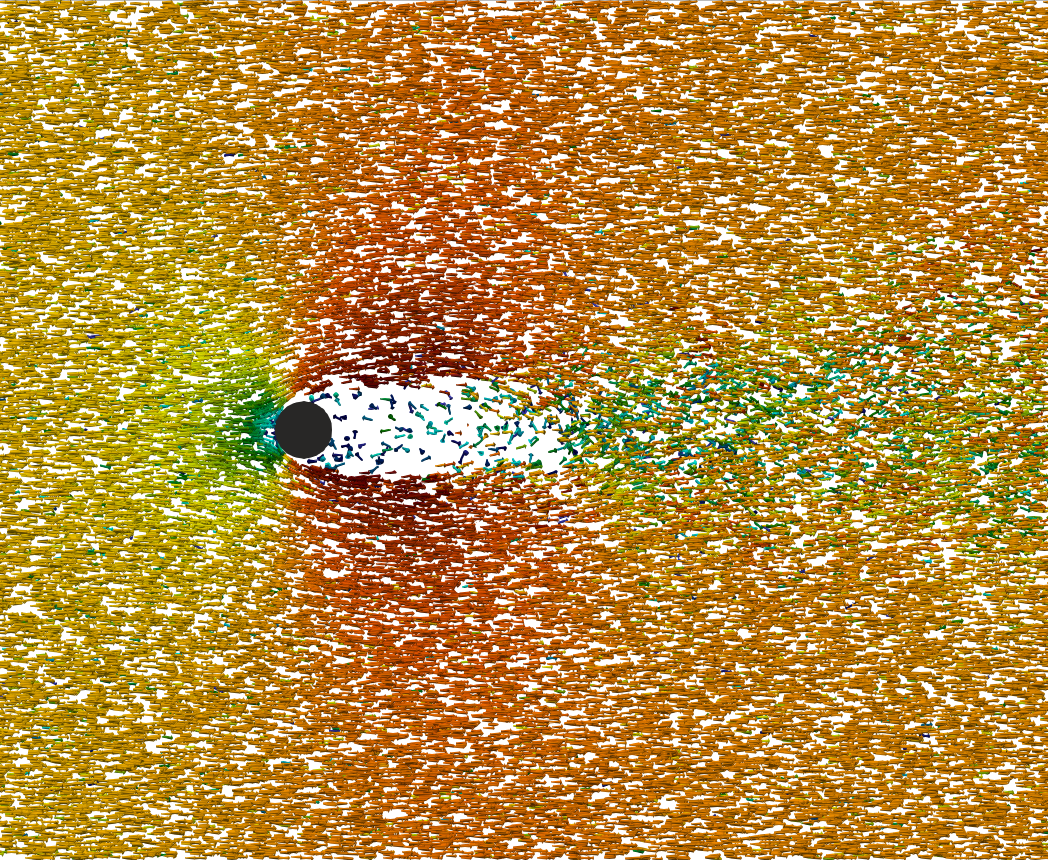}
  \caption{$St = \num{1.0}$}
  \label{fig:application:cylinder:dist:st1}
  \end{subfigure}\hfill
  \begin{subfigure}[b]{.29\textwidth}
  \includegraphics[width=\textwidth]{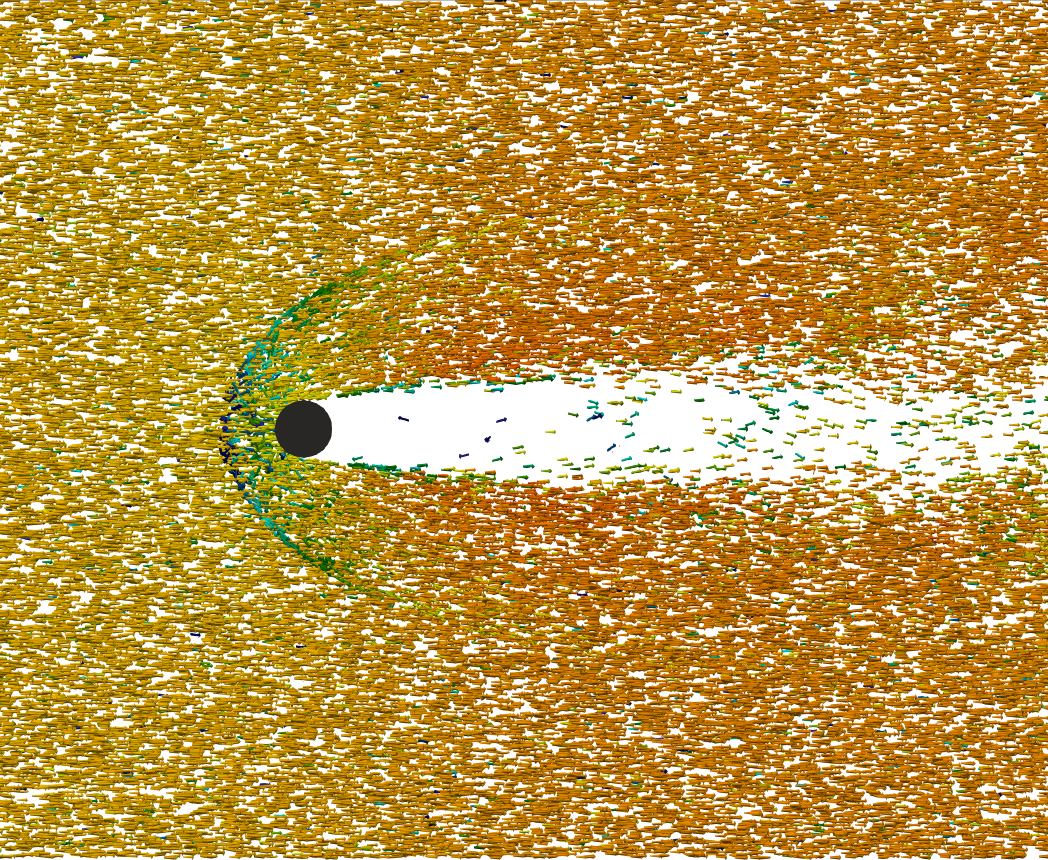}
  \caption{$St = \num{10.0}$}
  \label{fig:application:cylinder:dist:st10}
  \end{subfigure}\hfill
  \begin{subfigure}[b]{.0625\textwidth}
  \includegraphics[width=\textwidth]{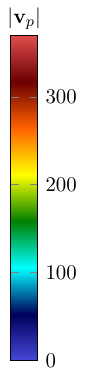}
  \end{subfigure}
  \caption{Particle distribution at $z \leq H$ for different Stokes numbers colored by velocity magnitude $\abs{\partvel}$.}
  \label{fig:application:cylinder:dist}
\end{figure}

Particles were emitted at the end of the RRALF region with the particle velocity chosen as the instantaneous fluid velocity.
The particle density is set to $\partdens = \SI{2500}{kg/m^3}$.
A total of \num{11} particle species were simulated with Stokes numbers ranging from \numrange{0.001}{10.}.
The instantaneous particle position was sampled every \SI{8.3e-08}{\s}, with the cumulative particle distribution at height $z \leq H$ depicted in~\cref{fig:application:cylinder:dist}.
Lower ($St = 0.1$,~\cref{fig:application:cylinder:dist:st01}) and higher ($St=10$,~\cref{fig:application:cylinder:dist:st10}) Stokes
numbers result in a more pronounced particle-free region compared to the medium Stokes number case
in~\cref{fig:application:cylinder:dist:st1} which shows the fastest return to (almost) uniform distribution, albeit with a strong
variation in particle momentum. Especially the high Stokes number case yields a local increase in particle density with the impact extending several times the diameter of the cylinder downstream.
Here, the working high-order boundary treatment is directly evident in~\cref{fig:application:cylinder:dist:st10}, as particles reflect off the curved surface and move upstream in the laminar flow outside the wall boundary layer until the downstream force causes a reversal of the velocity vector.
The resulting bimodal distribution between reflected and non-reflected particles needs to be considered when predicting compressor erosion, particularly of the subsequent stages.

\section{Conclusion and Outlook}
\label{sec:conclusion}
Particle-laden flows pose significant challenges towards their efficient and accurate numerical solution, especially in a high performance setting.
In the context of the compressible Navier-Stokes equations, the focus of one- or two-way coupled Euler-Lagrangian solvers is more on the time-accurate
particle tracking than on the efficiency on highly parallel systems.
In this work, we aimed to alleviate this deficiency and presented the extension of the open-source massively parallel solver FLEXI towards particle-laden flows.
Since the Eulerian code base was previously validated, this work focused the modeling of the dispersed phase and its numerical treatment.
Particular emphasis was given to the implementation of the particle tracking on parallel systems with arbitrary and possibly curved element faces.
Since extensibility was an explicit goal, we chose to extensively discuss the background and motivation leading to specific implementation choices.
Subsequently, each building block of our proposed framework was validated on its own, and we have verified the predictive performance of our implementation.
Furthermore, we illustrated the excellent scaling properties of our framework using a canonical test case.
Finally, the complete framework was applied to more challenging test cases to demonstrate its applicability to large-scale problems.
In the future, we seek to extend this open-source framework to incorporate further features such as the tracking of
larger particles with a level-set ansatz or the extension to four-way coupling. At the same time, our project is open towards any external contribution.

\section*{Acknowledgements}
The research presented in this paper was funded in parts by Deutsche Forschungsgemeinschaft (DFG, German Research
Foundation) under Germany's Excellence Strategy - EXC 2075 - 390740016 and by the DFG Rebound - 420603919.
We acknowledge the support by the Stuttgart Center for Simulation Science (SimTech).
The authors gratefully acknowledge the support and the computing time on “Hawk” provided by the HLRS through the project “hpcdg”.

\bibliographystyle{elsarticle-num-names}
\bibliography{references}

\begin{thebibliography}{100}
\expandafter\ifx\csname natexlab\endcsname\relax\def\natexlab#1{#1}\fi
\providecommand{\url}[1]{\texttt{#1}}
\providecommand{\href}[2]{#2}
\providecommand{\path}[1]{#1}
\providecommand{\DOIprefix}{doi:}
\providecommand{\ArXivprefix}{arXiv:}
\providecommand{\URLprefix}{URL: }
\providecommand{\Pubmedprefix}{pmid:}
\providecommand{\doi}[1]{\href{http://dx.doi.org/#1}{\path{#1}}}
\providecommand{\Pubmed}[1]{\href{pmid:#1}{\path{#1}}}
\providecommand{\bibinfo}[2]{#2}
\ifx\xfnm\relax \def\xfnm[#1]{\unskip,\space#1}\fi
\bibitem[{Delannay et~al.(2017)Delannay, Valance, Mangeney, Roche, and
  Richard}]{Delannay2017}
\bibinfo{author}{R.~Delannay}, \bibinfo{author}{A.~Valance},
  \bibinfo{author}{A.~Mangeney}, \bibinfo{author}{O.~Roche},
  \bibinfo{author}{P.~Richard},
\newblock \bibinfo{title}{Granular and particle-laden flows: from laboratory
  experiments to field observations},
\newblock \bibinfo{journal}{Journal of Physics D: Applied Physics}
  \bibinfo{volume}{50} (\bibinfo{year}{2017}) \bibinfo{pages}{053001}.
  \DOIprefix\doi{10.1088/1361-6463/50/5/053001}.
\bibitem[{Brandt and Coletti(2022)}]{Brandt2022}
\bibinfo{author}{L.~Brandt}, \bibinfo{author}{F.~Coletti},
\newblock \bibinfo{title}{Particle-laden turbulence: Progress and
  perspectives},
\newblock \bibinfo{journal}{Annual Review of Fluid Mechanics}
  \bibinfo{volume}{54} (\bibinfo{year}{2022}) \bibinfo{pages}{159--189}.
  \DOIprefix\doi{10.1146/annurev-fluid-030121-021103}.
\bibitem[{Higson et~al.(1994)Higson, Griffiths, Jones, and Hall}]{Higson1994}
\bibinfo{author}{H.~L. Higson}, \bibinfo{author}{R.~Griffiths},
  \bibinfo{author}{C.~Jones}, \bibinfo{author}{D.~J. Hall},
\newblock \bibinfo{title}{Concentration measurements around an isolated
  building: A comparison between wind tunnel and field data},
\newblock \bibinfo{journal}{Atmospheric Environment} \bibinfo{volume}{28}
  (\bibinfo{year}{1994}) \bibinfo{pages}{1827--1836}.
  \DOIprefix\doi{10.1016/1352-2310(94)90322-0}.
\bibitem[{Fernando and Choi(2007)}]{Fernando2007}
\bibinfo{author}{H.~J.~S. Fernando}, \bibinfo{author}{Y.-J. Choi},
\newblock \bibinfo{title}{Particle laden geophysical flows: from geophysical to
  sub-kolmogorov scales},
\newblock in: \bibinfo{booktitle}{{ERCOFTAC} Series},
  \bibinfo{publisher}{Springer Netherlands}, \bibinfo{year}{2007}, pp.
  \bibinfo{pages}{407--421}. \DOIprefix\doi{10.1007/978-1-4020-6218-6\_32}.
\bibitem[{Hefny and Ooka(2009)}]{Hefny2009}
\bibinfo{author}{M.~M. Hefny}, \bibinfo{author}{R.~Ooka},
\newblock \bibinfo{title}{{CFD} analysis of pollutant dispersion around
  buildings: Effect of cell geometry},
\newblock \bibinfo{journal}{Building and Environment} \bibinfo{volume}{44}
  (\bibinfo{year}{2009}) \bibinfo{pages}{1699--1706}.
  \DOIprefix\doi{10.1016/j.buildenv.2008.11.010}.
\bibitem[{Chang et~al.(2006)Chang, Hsieh, and Kao}]{Chang2006}
\bibinfo{author}{T.-J. Chang}, \bibinfo{author}{Y.-F. Hsieh},
  \bibinfo{author}{H.-M. Kao},
\newblock \bibinfo{title}{Numerical investigation of airflow pattern and
  particulate matter transport in naturally ventilated multi-room buildings},
\newblock \bibinfo{journal}{Indoor Air} \bibinfo{volume}{16}
  (\bibinfo{year}{2006}) \bibinfo{pages}{136--152}.
  \DOIprefix\doi{10.1111/j.1600-0668.2005.00410.x}.
\bibitem[{Domino(2021)}]{Domino2021}
\bibinfo{author}{S.~P. Domino},
\newblock \bibinfo{title}{A case study on pathogen transport, deposition,
  evaporation and transmission: Linking high-fidelity computational fluid
  dynamics simulations to probability of infection},
\newblock \bibinfo{journal}{International Journal of Computational Fluid
  Dynamics} \bibinfo{volume}{35} (\bibinfo{year}{2021})
  \bibinfo{pages}{743--757}. \DOIprefix\doi{10.1080/10618562.2021.1905801}.
\bibitem[{Ghenaiet(2012)}]{Ghenaiet2012a}
\bibinfo{author}{A.~Ghenaiet},
\newblock \bibinfo{title}{Study of sand particle trajectories and erosion into
  the first compression stage of a turbofan},
\newblock \bibinfo{journal}{Journal of Turbomachinery} \bibinfo{volume}{134}
  (\bibinfo{year}{2012}). \DOIprefix\doi{10.1115/1.4004750}.
\bibitem[{Marx et~al.(2014)Marx, St{\"{a}}ding, Reitz, and
  Friedrichs}]{Marx2014}
\bibinfo{author}{J.~Marx}, \bibinfo{author}{J.~St{\"{a}}ding},
  \bibinfo{author}{G.~Reitz}, \bibinfo{author}{J.~Friedrichs},
\newblock \bibinfo{title}{Investigation and analysis of deterioration in high
  pressure compressors due to operation},
\newblock \bibinfo{journal}{{CEAS} Aeronautical Journal} \bibinfo{volume}{5}
  (\bibinfo{year}{2014}) \bibinfo{pages}{515--525}.
  \DOIprefix\doi{10.1007/s13272-014-0118-z}.
\bibitem[{Sommerfeld et~al.(2021)Sommerfeld, Koch, Schwarz, and
  Beck}]{Sommerfeld2021}
\bibinfo{author}{H.~Sommerfeld}, \bibinfo{author}{C.~Koch},
  \bibinfo{author}{A.~Schwarz}, \bibinfo{author}{A.~Beck},
\newblock \bibinfo{title}{High velocity measurements of particle rebound
  characteristics under erosive conditions of high pressure compressors},
\newblock \bibinfo{journal}{Wear} \bibinfo{volume}{470-471}
  (\bibinfo{year}{2021}) \bibinfo{pages}{203626}.
  \DOIprefix\doi{10.1016/j.wear.2021.203626}.
\bibitem[{McDonald and Menon(2005)}]{McDonald2005}
\bibinfo{author}{B.~A. McDonald}, \bibinfo{author}{S.~Menon},
\newblock \bibinfo{title}{Direct numerical simulation of solid propellant
  combustion in crossflow},
\newblock \bibinfo{journal}{Journal of Propulsion and Power}
  \bibinfo{volume}{21} (\bibinfo{year}{2005}) \bibinfo{pages}{460--469}.
  \DOIprefix\doi{10.2514/1.10049}.
\bibitem[{Jones et~al.(2014)Jones, Marquis, and Vogiatzaki}]{Jones2014}
\bibinfo{author}{W.~Jones}, \bibinfo{author}{A.~Marquis},
  \bibinfo{author}{K.~Vogiatzaki},
\newblock \bibinfo{title}{Large-eddy simulation of spray combustion in a gas
  turbine combustor},
\newblock \bibinfo{journal}{Combustion and Flame} \bibinfo{volume}{161}
  (\bibinfo{year}{2014}) \bibinfo{pages}{222--239}.
  \DOIprefix\doi{10.1016/j.combustflame.2013.07.016}.
\bibitem[{Elghobashi(1994)}]{Elghobashi1994}
\bibinfo{author}{S.~Elghobashi},
\newblock \bibinfo{title}{On predicting particle-laden turbulent flows},
\newblock \bibinfo{journal}{Applied Scientific Research} \bibinfo{volume}{52}
  (\bibinfo{year}{1994}) \bibinfo{pages}{309--329}.
  \DOIprefix\doi{10.1007/BF00936835}.
\bibitem[{Vance and Squires(2002)}]{Vance2002}
\bibinfo{author}{M.~W. Vance}, \bibinfo{author}{K.~D. Squires},
\newblock \bibinfo{title}{An approach to parallel computing in an
  eulerian-lagrangian two-phase flow model},
\newblock in: \bibinfo{booktitle}{Volume 2: Symposia and General Papers, Parts
  A and B}, \bibinfo{publisher}{ASMEDC}, \bibinfo{year}{2002}.
  \DOIprefix\doi{10.1115/fedsm2002-31225}.
\bibitem[{Balachandar and Eaton(2010)}]{Balachandar2010}
\bibinfo{author}{S.~Balachandar}, \bibinfo{author}{J.~K. Eaton},
\newblock \bibinfo{title}{Turbulent dispersed multiphase flow},
\newblock \bibinfo{journal}{Annual Review of Fluid Mechanics}
  \bibinfo{volume}{42} (\bibinfo{year}{2010}) \bibinfo{pages}{111--133}.
  \DOIprefix\doi{10.1146/annurev.fluid.010908.165243}.
\bibitem[{Kuerten(2016)}]{Kuerten2016}
\bibinfo{author}{J.~G.~M. Kuerten},
\newblock \bibinfo{title}{Point-particle {DNS} and {LES} of particle-laden
  turbulent flow - a state-of-the-art review},
\newblock \bibinfo{journal}{Flow, Turbulence and Combustion}
  \bibinfo{volume}{97} (\bibinfo{year}{2016}) \bibinfo{pages}{689--713}.
  \DOIprefix\doi{10.1007/s10494-016-9765-y}.
\bibitem[{Beck et~al.(2019)Beck, Ortwein, Kopper, Krais, Kempf, and
  Koch}]{Beck2019}
\bibinfo{author}{A.~Beck}, \bibinfo{author}{P.~Ortwein},
  \bibinfo{author}{P.~Kopper}, \bibinfo{author}{N.~Krais},
  \bibinfo{author}{D.~Kempf}, \bibinfo{author}{C.~Koch},
\newblock \bibinfo{title}{Towards high-fidelity erosion prediction: On
  time-accurate particle tracking in turbomachinery},
\newblock \bibinfo{journal}{Int J Heat Fluid Flow} \bibinfo{volume}{79}
  (\bibinfo{year}{2019}) \bibinfo{pages}{108457}.
\bibitem[{Kopper et~al.(2021)Kopper, Kurz, Wenzel, D{\"{u}}rrw{\"{a}}chter,
  Koch, and Beck}]{Kopper2021a}
\bibinfo{author}{P.~Kopper}, \bibinfo{author}{M.~Kurz},
  \bibinfo{author}{C.~Wenzel}, \bibinfo{author}{J.~D{\"{u}}rrw{\"{a}}chter},
  \bibinfo{author}{C.~Koch}, \bibinfo{author}{A.~Beck},
\newblock \bibinfo{title}{Boundary-layer dynamics in wall-resolved {LES} across
  multiple turbine stages}  (\bibinfo{year}{2021}) \bibinfo{pages}{1--14}.
  \DOIprefix\doi{10.2514/1.j060633}.
\bibitem[{Kaiser et~al.(2021)Kaiser, Appel, Fritz, Adami, and
  Adams}]{Kaiser2021}
\bibinfo{author}{J.~Kaiser}, \bibinfo{author}{D.~Appel},
  \bibinfo{author}{F.~Fritz}, \bibinfo{author}{S.~Adami},
  \bibinfo{author}{N.~Adams},
\newblock \bibinfo{title}{A multiresolution local-timestepping scheme for
  particle-laden multiphase flow simulations using a level-set and
  point-particle approach},
\newblock \bibinfo{journal}{Computer Methods in Applied Mechanics and
  Engineering} \bibinfo{volume}{384} (\bibinfo{year}{2021})
  \bibinfo{pages}{113966}. \DOIprefix\doi{10.1016/j.cma.2021.113966}.
\bibitem[{Patel and Capecelatro(2022)}]{Patel2022}
\bibinfo{author}{M.~Patel}, \bibinfo{author}{J.~Capecelatro},
\newblock \bibinfo{title}{A high-order low-dissipation euler-lagrange method
  for compressible gas-particle flows},
\newblock in: \bibinfo{booktitle}{11th International Conference on
  Computational Fluid Dynamics}, \bibinfo{year}{2022}, pp.
  \bibinfo{pages}{1--14}.
\bibitem[{Hoppe et~al.(2022)Hoppe, Winter, Adami, and Adams}]{Hoppe2022}
\bibinfo{author}{N.~Hoppe}, \bibinfo{author}{J.~M. Winter},
  \bibinfo{author}{S.~Adami}, \bibinfo{author}{N.~A. Adams},
\newblock \bibinfo{title}{{ALPACA} - a level-set based sharp-interface
  multiresolution solver for conservation laws},
\newblock \bibinfo{journal}{Computer Physics Communications}
  \bibinfo{volume}{272} (\bibinfo{year}{2022}) \bibinfo{pages}{108246}.
  \URLprefix \url{https://doi.org/10.1016/j.cpc.2021.108246}.
  \DOIprefix\doi{10.1016/j.cpc.2021.108246}.
\bibitem[{Hindenlang et~al.(2012)Hindenlang, Gassner, Altmann, Beck,
  Staudenmaier, and Munz}]{Hindenlang2012}
\bibinfo{author}{F.~Hindenlang}, \bibinfo{author}{G.~J. Gassner},
  \bibinfo{author}{C.~Altmann}, \bibinfo{author}{A.~Beck},
  \bibinfo{author}{M.~Staudenmaier}, \bibinfo{author}{C.-D. Munz},
\newblock \bibinfo{title}{Explicit discontinuous galerkin methods for unsteady
  problems},
\newblock \bibinfo{journal}{Computers {\&} Fluids} \bibinfo{volume}{61}
  (\bibinfo{year}{2012}) \bibinfo{pages}{86--93}.
  \DOIprefix\doi{10.1016/j.compfluid.2012.03.006}.
\bibitem[{Gassner and Beck(2012)}]{Gassner2012}
\bibinfo{author}{G.~J. Gassner}, \bibinfo{author}{A.~D. Beck},
\newblock \bibinfo{title}{On the accuracy of high-order discretizations for
  underresolved turbulence simulations},
\newblock \bibinfo{journal}{Theoretical and Computational Fluid Dynamics}
  \bibinfo{volume}{27} (\bibinfo{year}{2012}) \bibinfo{pages}{221--237}.
  \DOIprefix\doi{10.1007/s00162-011-0253-7}.
\bibitem[{Krais et~al.(2021)Krais, Beck, Bolemann, Frank, Flad, Gassner,
  Hindenlang, Hoffmann, Kuhn, Sonntag, and Munz}]{Krais2021}
\bibinfo{author}{N.~Krais}, \bibinfo{author}{A.~Beck},
  \bibinfo{author}{T.~Bolemann}, \bibinfo{author}{H.~Frank},
  \bibinfo{author}{D.~Flad}, \bibinfo{author}{G.~Gassner},
  \bibinfo{author}{F.~Hindenlang}, \bibinfo{author}{M.~Hoffmann},
  \bibinfo{author}{T.~Kuhn}, \bibinfo{author}{M.~Sonntag},
  \bibinfo{author}{C.-D. Munz},
\newblock \bibinfo{title}{{FLEXI}: A high order discontinuous {G}alerkin
  framework for hyperbolic{\textendash}parabolic conservation laws},
\newblock \bibinfo{journal}{Comput. Math. with Appl.} \bibinfo{volume}{81}
  (\bibinfo{year}{2021}) \bibinfo{pages}{186--219}.
\bibitem[{Beck et~al.(2018)Beck, Bolemann, Flad, Frank, Krais, Kukuschkin,
  Sonntag, and Munz}]{Beck2018}
\bibinfo{author}{A.~Beck}, \bibinfo{author}{T.~Bolemann},
  \bibinfo{author}{D.~Flad}, \bibinfo{author}{H.~Frank},
  \bibinfo{author}{N.~Krais}, \bibinfo{author}{K.~Kukuschkin},
  \bibinfo{author}{M.~Sonntag}, \bibinfo{author}{C.-D. Munz},
\newblock \bibinfo{title}{Application and development of the high order
  discontinuous galerkin spectral element method for compressible multiscale
  flows},
\newblock in: \bibinfo{booktitle}{High Performance Computing in Science and
  Engineering '17}, \bibinfo{publisher}{Springer International Publishing},
  \bibinfo{year}{2018}, pp. \bibinfo{pages}{387--407}.
  \DOIprefix\doi{10.1007/978-3-319-68394-2\_23}.
\bibitem[{Beck et~al.(2020)Beck, Zeifang, Schwarz, and Flad}]{Beck2020}
\bibinfo{author}{A.~D. Beck}, \bibinfo{author}{J.~Zeifang},
  \bibinfo{author}{A.~Schwarz}, \bibinfo{author}{D.~G. Flad},
\newblock \bibinfo{title}{A neural network based shock detection and
  localization approach for discontinuous {G}alerkin methods},
\newblock \bibinfo{journal}{Journal of Computational Physics}
  \bibinfo{volume}{423} (\bibinfo{year}{2020}) \bibinfo{pages}{109824}.
  \DOIprefix\doi{10.1016/j.jcp.2020.109824}.
\bibitem[{Zeifang and Beck(2021)}]{Zeifang2021}
\bibinfo{author}{J.~Zeifang}, \bibinfo{author}{A.~Beck},
\newblock \bibinfo{title}{A data-driven high order sub-cell artificial
  viscosity for the discontinuous galerkin spectral element method},
\newblock \bibinfo{journal}{Journal of Computational Physics}
  \bibinfo{volume}{441} (\bibinfo{year}{2021}) \bibinfo{pages}{110475}.
  \DOIprefix\doi{10.1016/j.jcp.2021.110475}.
\bibitem[{Sonntag(2017)}]{SonntagPHD}
\bibinfo{author}{M.~Sonntag}, \bibinfo{title}{Shape derivatives and shock
  capturing for the Navier-Stokes equations in discontinuous Galerkin methods},
  Ph.D. thesis, Universit{\"{a}}t Stuttgart, \bibinfo{year}{2017}.
  \DOIprefix\doi{10.18419/opus-9342}.
\bibitem[{Kurz et~al.(2022)Kurz, Offenh{\"{a}}user, Viola, Resch, and
  Beck}]{Kurz2022}
\bibinfo{author}{M.~Kurz}, \bibinfo{author}{P.~Offenh{\"{a}}user},
  \bibinfo{author}{D.~Viola}, \bibinfo{author}{M.~Resch},
  \bibinfo{author}{A.~Beck},
\newblock \bibinfo{title}{Relexi--a scalable open source reinforcement learning
  framework for high-performance computing},
\newblock \bibinfo{journal}{Software Impacts}  (\bibinfo{year}{2022})
  \bibinfo{pages}{100422}. \DOIprefix\doi{10.1016/j.simpa.2022.100422}.
\bibitem[{Sutherland(1893)}]{Sutherland1893}
\bibinfo{author}{W.~Sutherland},
\newblock \bibinfo{title}{Lii. the viscosity of gases and molecular force},
\newblock \bibinfo{journal}{The London, Edinburgh, and Dublin Philosophical
  Magazine and Journal of Science} \bibinfo{volume}{36} (\bibinfo{year}{1893})
  \bibinfo{pages}{507--531}. \DOIprefix\doi{10.1080/14786449308620508}.
\bibitem[{Basset(1888)}]{basset1888treatise}
\bibinfo{author}{A.~B. Basset}, \bibinfo{title}{A treatise on hydrodynamics:
  with numerous examples}, volume~\bibinfo{volume}{2},
  \bibinfo{publisher}{Deighton, Bell and Company}, \bibinfo{year}{1888}.
\bibitem[{Boussinesq(1885)}]{boussinesq1885resistance}
\bibinfo{author}{J.~Boussinesq},
\newblock \bibinfo{title}{Sur la resistance qu'oppose un fluide indefini en
  repos, sans pesanteur, au mouvement varie d'une sphere solide qu'il mouille
  sur toute sa surface, quand les vitesses restent bien continues et assez
  faibles pour que leurs carres et produits soient negligiables},
\newblock \bibinfo{journal}{C. R. Acad. Sci. Paris} \bibinfo{volume}{100}
  (\bibinfo{year}{1885}) \bibinfo{pages}{935--937}.
\bibitem[{Oseen(1927)}]{oseen1927neuere}
\bibinfo{author}{C.~W. Oseen},
\newblock \bibinfo{title}{Neuere {M}ethoden und {E}rgebnisse in der
  {H}ydrodynamik},
\newblock \bibinfo{journal}{Mathematik und ihre Anwendungen in Monographien und
  Lehrb\"{u}chern}  (\bibinfo{year}{1927}).
\bibitem[{Tchen(1947)}]{Tchen1947}
\bibinfo{author}{C.~Tchen}, \bibinfo{title}{Mean value and correlation problems
  connected with the motion of small particles suspended in a turbulent fluid},
  Ph.D. thesis, Delft University, \bibinfo{year}{1947}.
\bibitem[{Corrsin and Lumley(1956)}]{Corrsin1956}
\bibinfo{author}{S.~Corrsin}, \bibinfo{author}{J.~Lumley},
\newblock \bibinfo{title}{On the equation of motion for a particle in turbulent
  fluid},
\newblock \bibinfo{journal}{Applied Scientific Research} \bibinfo{volume}{6}
  (\bibinfo{year}{1956}) \bibinfo{pages}{114--116}.
  \DOIprefix\doi{10.1007/bf03185030}.
\bibitem[{Auton et~al.(1988)Auton, Hunt, and Prud'Homme}]{Auton1988}
\bibinfo{author}{T.~R. Auton}, \bibinfo{author}{J.~C.~R. Hunt},
  \bibinfo{author}{M.~Prud'Homme},
\newblock \bibinfo{title}{The force exerted on a body in inviscid unsteady
  non-uniform rotational flow},
\newblock \bibinfo{journal}{Journal of Fluid Mechanics} \bibinfo{volume}{197}
  (\bibinfo{year}{1988}) \bibinfo{pages}{241--257}.
  \DOIprefix\doi{10.1017/S0022112088003246}.
\bibitem[{Maxey and Riley(1983)}]{Maxey1983}
\bibinfo{author}{M.~R. Maxey}, \bibinfo{author}{J.~J. Riley},
\newblock \bibinfo{title}{Equation of motion for a small rigid sphere in a
  nonuniform flow},
\newblock \bibinfo{journal}{Physics of Fluids} \bibinfo{volume}{26}
  (\bibinfo{year}{1983}) \bibinfo{pages}{883--889}.
  \DOIprefix\doi{10.1063/1.864230}.
\bibitem[{Gatignol(1983)}]{Gatignol1983}
\bibinfo{author}{R.~Gatignol},
\newblock \bibinfo{title}{{The Fax{\'{e}}n formulae for a rigid particle in an
  unsteady non-uniform Stokes flow}},
\newblock \bibinfo{journal}{Journal de Mecanique Theorique et Appliquee}
  \bibinfo{volume}{2} (\bibinfo{year}{1983}) \bibinfo{pages}{143--160}.
\bibitem[{Fax{\'{e}}n(1922)}]{Faxen1922a}
\bibinfo{author}{H.~Fax{\'{e}}n},
\newblock \bibinfo{title}{{Der Widerstand gegen die Bewegung einer starren
  Kugel in einer z{\"{a}}hen Fl{\"{u}}ssigkeit, die zwischen zwei parallelen
  ebenen W{\"{a}}nden eingeschlossen ist}},
\newblock \bibinfo{journal}{Annalen der Physik} \bibinfo{volume}{373}
  (\bibinfo{year}{1922}) \bibinfo{pages}{89--119}.
  \DOIprefix\doi{10.1002/andp.19223731003}.
\bibitem[{Mei et~al.(1991)Mei, Adrian, and Hanratty}]{Mei1991}
\bibinfo{author}{R.~Mei}, \bibinfo{author}{R.~J. Adrian},
  \bibinfo{author}{T.~J. Hanratty},
\newblock \bibinfo{title}{Particle dispersion in isotropic turbulence under
  stokes drag and basset force with gravitational settling},
\newblock \bibinfo{journal}{Journal of Fluid Mechanics} \bibinfo{volume}{225}
  (\bibinfo{year}{1991}) \bibinfo{pages}{481--495}.
  \DOIprefix\doi{10.1017/S0022112091002136}.
\bibitem[{Basset(1889)}]{Basset1888}
\bibinfo{author}{A.~B. Basset},
\newblock \bibinfo{title}{Treatise on hydrodynamics},
\newblock \bibinfo{journal}{Nature} \bibinfo{volume}{40} (\bibinfo{year}{1889})
  \bibinfo{pages}{412--413}. \DOIprefix\doi{10.1038/040412a0}.
\bibitem[{Tatom(1988)}]{Tatom1988}
\bibinfo{author}{F.~B. Tatom},
\newblock \bibinfo{title}{The basset term as a semiderivative},
\newblock \bibinfo{journal}{Applied Scientific Research} \bibinfo{volume}{45}
  (\bibinfo{year}{1988}) \bibinfo{pages}{283--285}.
  \DOIprefix\doi{10.1007/BF00384691}.
\bibitem[{Farazmand and Haller(2015)}]{Farazmand2015}
\bibinfo{author}{M.~Farazmand}, \bibinfo{author}{G.~Haller},
\newblock \bibinfo{title}{The {M}axey-{R}iley equation: Existence, uniqueness
  and regularity of solutions},
\newblock \bibinfo{journal}{Nonlinear Analysis: Real World Applications}
  \bibinfo{volume}{22} (\bibinfo{year}{2015}) \bibinfo{pages}{98--106}.
  \DOIprefix\doi{10.1016/j.nonrwa.2014.08.002}.
\bibitem[{Crowe et~al.(2011)Crowe, Schwarzkopf, Sommerfeld, and
  Tsuji}]{Crowe2011}
\bibinfo{author}{C.~T. Crowe}, \bibinfo{author}{J.~D. Schwarzkopf},
  \bibinfo{author}{M.~Sommerfeld}, \bibinfo{author}{Y.~Tsuji},
  \bibinfo{title}{Multiphase Flows with Droplets and Particles},
  \bibinfo{publisher}{CRC Press}, \bibinfo{year}{2011}.
  \DOIprefix\doi{10.1201/b11103}.
\bibitem[{Parmar et~al.(2012)Parmar, Haselbacher, and Balachandar}]{Parmar2012}
\bibinfo{author}{M.~Parmar}, \bibinfo{author}{A.~Haselbacher},
  \bibinfo{author}{S.~Balachandar},
\newblock \bibinfo{title}{Equation of motion for a sphere in non-uniform
  compressible flows},
\newblock \bibinfo{journal}{Journal of Fluid Mechanics} \bibinfo{volume}{699}
  (\bibinfo{year}{2012}) \bibinfo{pages}{352--375}.
  \DOIprefix\doi{10.1017/jfm.2012.109}.
\bibitem[{Minier and Peirano(2001)}]{Minier2001}
\bibinfo{author}{J.-P. Minier}, \bibinfo{author}{E.~Peirano},
\newblock \bibinfo{title}{The pdf approach to turbulent polydispersed two-phase
  flows},
\newblock \bibinfo{journal}{Physics Reports} \bibinfo{volume}{352}
  (\bibinfo{year}{2001}) \bibinfo{pages}{1--214}.
  \DOIprefix\doi{10.1016/s0370-1573(01)00011-4}.
\bibitem[{Amiri et~al.(2006)Amiri, Hannani, and Mashayek}]{Amiri2006}
\bibinfo{author}{A.~E. Amiri}, \bibinfo{author}{S.~K. Hannani},
  \bibinfo{author}{F.~Mashayek},
\newblock \bibinfo{title}{Large-eddy simulation of heavy-particle transport in
  turbulent channel flow},
\newblock \bibinfo{journal}{Numerical Heat Transfer, Part B: Fundamentals}
  \bibinfo{volume}{50} (\bibinfo{year}{2006}) \bibinfo{pages}{285--313}.
  \DOIprefix\doi{10.1080/10407790600859577}.
\bibitem[{Breuer and Hoppe(2017)}]{Breuer2017}
\bibinfo{author}{M.~Breuer}, \bibinfo{author}{F.~Hoppe},
\newblock \bibinfo{title}{Influence of a cost{\textendash}efficient {Langevin}
  subgrid-scale model on the dispersed phase of large{\textendash}eddy
  simulations of turbulent bubble{\textendash}laden and
  particle{\textendash}laden flows},
\newblock \bibinfo{journal}{International Journal of Multiphase Flow}
  \bibinfo{volume}{89} (\bibinfo{year}{2017}) \bibinfo{pages}{23--44}.
  \DOIprefix\doi{10.1016/j.ijmultiphaseflow.2016.10.007}.
\bibitem[{Tropea et~al.(2007)Tropea, Yarin, Foss, and (Firm)}]{Tropea2007}
\bibinfo{author}{C.~Tropea}, \bibinfo{author}{A.~L. Yarin},
  \bibinfo{author}{J.~F. Foss}, \bibinfo{author}{T.~G. (Firm)},
  \bibinfo{title}{Springer Handbook of Experimental Fluid Mechanics},
  \bibinfo{publisher}{Springer}, \bibinfo{address}{Berlin},
  \bibinfo{year}{2007}.
\bibitem[{Schiller and Naumann(1933)}]{schiller1933grundlegenden}
\bibinfo{author}{L.~Schiller}, \bibinfo{author}{A.~Naumann},
\newblock \bibinfo{title}{{\"U}ber die grundlegenden {B}erechnungen bei der
  {S}chwerkraftaufbereitung},
\newblock \bibinfo{journal}{Z. Ver. Dtsch. Ing.} \bibinfo{volume}{77}
  (\bibinfo{year}{1933}) \bibinfo{pages}{318--320}.
\bibitem[{Haider and Levenspiel(1989)}]{Haider1989}
\bibinfo{author}{A.~Haider}, \bibinfo{author}{O.~Levenspiel},
\newblock \bibinfo{title}{Drag coefficient and terminal velocity of spherical
  and nonspherical particles},
\newblock \bibinfo{journal}{Powder Technology} \bibinfo{volume}{58}
  (\bibinfo{year}{1989}) \bibinfo{pages}{63--70}.
  \DOIprefix\doi{10.1016/0032-5910(89)80008-7}.
\bibitem[{Loth(2008)}]{Loth2008a}
\bibinfo{author}{E.~Loth},
\newblock \bibinfo{title}{Compressibility and rarefaction effects on drag of a
  spherical particle},
\newblock \bibinfo{journal}{AIAA Journal} \bibinfo{volume}{46}
  (\bibinfo{year}{2008}) \bibinfo{pages}{2219--2228}. \URLprefix
  \url{https://arc.aiaa.org/doi/10.2514/1.28943}.
  \DOIprefix\doi{10.2514/1.28943}.
\bibitem[{Saffman(1965)}]{Saffman1965}
\bibinfo{author}{P.~G. Saffman},
\newblock \bibinfo{title}{The lift on a small sphere in a slow shear flow},
\newblock \bibinfo{journal}{Journal of Fluid Mechanics} \bibinfo{volume}{22}
  (\bibinfo{year}{1965}) \bibinfo{pages}{385--400}.
  \DOIprefix\doi{10.1017/S0022112065000824}.
\bibitem[{Saffman(1968)}]{Saffman1968}
\bibinfo{author}{P.~G. Saffman},
\newblock \bibinfo{title}{The lift on a small sphere in a slow shear flow -
  corrigendum},
\newblock \bibinfo{journal}{Journal of Fluid Mechanics} \bibinfo{volume}{31}
  (\bibinfo{year}{1968}) \bibinfo{pages}{624--624}.
  \DOIprefix\doi{10.1017/S0022112068999990}.
\bibitem[{Mei(1992)}]{Mei1992}
\bibinfo{author}{R.~Mei},
\newblock \bibinfo{title}{An approximate expression for the shear lift force on
  a spherical particle at finite reynolds number},
\newblock \bibinfo{journal}{International Journal of Multiphase Flow}
  \bibinfo{volume}{18} (\bibinfo{year}{1992}) \bibinfo{pages}{145--147}.
  \DOIprefix\doi{10.1016/0301-9322(92)90012-6}.
\bibitem[{Rubinow and Keller(1961)}]{Rubinow1961}
\bibinfo{author}{S.~I. Rubinow}, \bibinfo{author}{J.~B. Keller},
\newblock \bibinfo{title}{The transverse force on a spinning sphere moving in a
  viscous fluid},
\newblock \bibinfo{journal}{Journal of Fluid Mechanics} \bibinfo{volume}{11}
  (\bibinfo{year}{1961}) \bibinfo{pages}{447}.
  \DOIprefix\doi{10.1017/S0022112061000640}.
\bibitem[{Oesterl{\'{e}} and Dinh(1998)}]{Oesterle1998}
\bibinfo{author}{B.~Oesterl{\'{e}}}, \bibinfo{author}{T.~B. Dinh},
\newblock \bibinfo{title}{Experiments on the lift of a spinning sphere in a
  range of intermediate reynolds numbers},
\newblock \bibinfo{journal}{Experiments in Fluids} \bibinfo{volume}{25}
  (\bibinfo{year}{1998}) \bibinfo{pages}{16--22}.
  \DOIprefix\doi{10.1007/s003480050203}.
\bibitem[{Feuillebois and Lasek(1978)}]{Feuillebois1978}
\bibinfo{author}{F.~Feuillebois}, \bibinfo{author}{A.~Lasek},
\newblock \bibinfo{title}{Significant degeneracies of the equations of a
  suspension for large reynolds number},
\newblock in: \bibinfo{booktitle}{Numerical Methods in Laminar and Turbulent
  Flow}, \bibinfo{year}{1978}, pp. \bibinfo{pages}{171--177}.
\bibitem[{Dennis et~al.(1980)Dennis, Singh, and Ingham}]{Dennis1980}
\bibinfo{author}{S.~C.~R. Dennis}, \bibinfo{author}{S.~N. Singh},
  \bibinfo{author}{D.~B. Ingham},
\newblock \bibinfo{title}{The steady flow due to a rotating sphere at low and
  moderate reynolds numbers},
\newblock \bibinfo{journal}{Journal of Fluid Mechanics} \bibinfo{volume}{101}
  (\bibinfo{year}{1980}) \bibinfo{pages}{257--279}.
  \DOIprefix\doi{10.1017/S0022112080001656}.
\bibitem[{Reeks(1983)}]{Reeks1983}
\bibinfo{author}{M.~Reeks},
\newblock \bibinfo{title}{The transport of discrete particles in inhomogeneous
  turbulence},
\newblock \bibinfo{journal}{Journal of Aerosol Science} \bibinfo{volume}{14}
  (\bibinfo{year}{1983}) \bibinfo{pages}{729--739}.
  \DOIprefix\doi{10.1016/0021-8502(83)90055-1}.
\bibitem[{van Hinsberg et~al.(2011)van Hinsberg, {ten Thije Boonkkamp}, and
  Clercx}]{VanHinsberg2011}
\bibinfo{author}{M.~van Hinsberg}, \bibinfo{author}{J.~{ten Thije Boonkkamp}},
  \bibinfo{author}{H.~Clercx},
\newblock \bibinfo{title}{An efficient, second order method for the
  approximation of the basset history force},
\newblock \bibinfo{journal}{Journal of Computational Physics}
  \bibinfo{volume}{230} (\bibinfo{year}{2011}) \bibinfo{pages}{1465--1478}.
  \DOIprefix\doi{10.1016/j.jcp.2010.11.014}.
  \href{http://arxiv.org/abs/1008.0833}{{\tt arXiv:1008.0833}}.
\bibitem[{Crowe et~al.(1977)Crowe, Sharma, and Stock}]{Crowe1977}
\bibinfo{author}{C.~T. Crowe}, \bibinfo{author}{M.~P. Sharma},
  \bibinfo{author}{D.~E. Stock},
\newblock \bibinfo{title}{The particle-source-in cell (psi-cell) model for
  gas-droplet flows},
\newblock \bibinfo{journal}{Journal of Fluids Engineering} \bibinfo{volume}{99}
  (\bibinfo{year}{1977}) \bibinfo{pages}{325--332}.
  \DOIprefix\doi{10.1115/1.3448756}.
\bibitem[{Horwitz and Mani(2016)}]{Horwitz2016b}
\bibinfo{author}{J.~Horwitz}, \bibinfo{author}{A.~Mani},
\newblock \bibinfo{title}{Accurate calculation of stokes drag for
  point-particle tracking in two-way coupled flows},
\newblock \bibinfo{journal}{Journal of Computational Physics}
  \bibinfo{volume}{318} (\bibinfo{year}{2016}) \bibinfo{pages}{85--109}.
  \DOIprefix\doi{10.1016/j.jcp.2016.04.034}.
\bibitem[{Tabakoff and Wakeman(1981)}]{Tabakoff1981}
\bibinfo{author}{W.~Tabakoff}, \bibinfo{author}{T.~Wakeman},
  \bibinfo{title}{Basic Erosion Investigation in Small Turbomachinery},
  \bibinfo{type}{Technical Report}, Cincinnati University, Dept. of Aerospace
  Engineering and Applied Mechanics, \bibinfo{year}{1981}.
\bibitem[{Bons et~al.(2017)Bons, Prenter, and Whitaker}]{Bons2017}
\bibinfo{author}{J.~P. Bons}, \bibinfo{author}{R.~Prenter},
  \bibinfo{author}{S.~Whitaker},
\newblock \bibinfo{title}{A simple physics-based model for particle rebound and
  deposition in turbomachinery},
\newblock \bibinfo{journal}{Journal of Turbomachinery} \bibinfo{volume}{139}
  (\bibinfo{year}{2017}) \bibinfo{pages}{081009}.
  \DOIprefix\doi{10.1115/1.4035921}.
\bibitem[{Whitaker and Bons(2018)}]{Whitaker2018}
\bibinfo{author}{S.~M. Whitaker}, \bibinfo{author}{J.~P. Bons},
\newblock \bibinfo{title}{An improved particle impact model by accounting for
  rate of strain and stochastic rebound},
\newblock in: \bibinfo{booktitle}{Turbo Expo: Power for Land, Sea, and Air},
  volume \bibinfo{volume}{Volume 2D: Turbomachinery}, \bibinfo{year}{2018}.
  \DOIprefix\doi{10.1115/GT2018-77158}, \bibinfo{note}{v02DT47A016}.
\bibitem[{Schwarz et~al.(2022)Schwarz, Kopper, Keim, Sommerfeld, Koch, and
  Beck}]{Schwarz2022}
\bibinfo{author}{A.~Schwarz}, \bibinfo{author}{P.~Kopper},
  \bibinfo{author}{J.~Keim}, \bibinfo{author}{H.~Sommerfeld},
  \bibinfo{author}{C.~Koch}, \bibinfo{author}{A.~Beck},
\newblock \bibinfo{title}{A neural network based framework to model particle
  rebound and fracture},
\newblock \bibinfo{journal}{Wear}  (\bibinfo{year}{2022})
  \bibinfo{pages}{204476}. \DOIprefix\doi{10.1016/j.wear.2022.204476}.
\bibitem[{Hindenlang et~al.(2015)Hindenlang, Bolemann, and
  Munz}]{Hindenlang2015}
\bibinfo{author}{F.~Hindenlang}, \bibinfo{author}{T.~Bolemann},
  \bibinfo{author}{C.~D. Munz}, \bibinfo{title}{Mesh Curving Techniques for
  High Order {D}iscontinuous {G}alerkin Simulations},
  \bibinfo{publisher}{Springer International Publishing},
  \bibinfo{address}{Cham}, \bibinfo{year}{2015}, pp. \bibinfo{pages}{133--152}.
  \DOIprefix\doi{10.1007/978-3-319-12886-3\_8}.
\bibitem[{Toro(2009)}]{Toro2009}
\bibinfo{author}{E.~F. Toro}, \bibinfo{title}{Riemann Solvers and Numerical
  Methods for Fluid Dynamics}, \bibinfo{publisher}{Springer Berlin Heidelberg},
  \bibinfo{address}{Berlin, Heidelberg}, \bibinfo{year}{2009}.
  \DOIprefix\doi{10.1007/b79761}.
\bibitem[{Harten and Hyman(1983)}]{Harten1983b}
\bibinfo{author}{A.~Harten}, \bibinfo{author}{J.~M. Hyman},
\newblock \bibinfo{title}{Self adjusting grid methods for one-dimensional
  hyperbolic conservation laws},
\newblock \bibinfo{journal}{Journal of Computational Physics}
  \bibinfo{volume}{50} (\bibinfo{year}{1983}) \bibinfo{pages}{235--269}.
  \DOIprefix\doi{10.1016/0021-9991(83)90066-9}.
\bibitem[{Bassi and Rebay(1997)}]{Bassi1997}
\bibinfo{author}{F.~Bassi}, \bibinfo{author}{S.~Rebay},
\newblock \bibinfo{title}{A high-order accurate discontinuous finite element
  method for the numerical solution of the compressible navier-stokes
  equations},
\newblock \bibinfo{journal}{Journal of Computational Physics}
  \bibinfo{volume}{131} (\bibinfo{year}{1997}) \bibinfo{pages}{267--279}.
  \DOIprefix\doi{10.1006/jcph.1996.5572}.
\bibitem[{Pirozzoli(2011)}]{Pirozzoli2011}
\bibinfo{author}{S.~Pirozzoli},
\newblock \bibinfo{title}{Numerical methods for high-speed flows},
\newblock \bibinfo{journal}{Annual Review of Fluid Mechanics}
  \bibinfo{volume}{43} (\bibinfo{year}{2011}) \bibinfo{pages}{163--194}.
  \DOIprefix\doi{10.1146/annurev-fluid-122109-160718}.
\bibitem[{Gassner et~al.(2016)Gassner, Winters, and Kopriva}]{Gassner2016}
\bibinfo{author}{G.~J. Gassner}, \bibinfo{author}{A.~R. Winters},
  \bibinfo{author}{D.~A. Kopriva},
\newblock \bibinfo{title}{Split form nodal discontinuous galerkin schemes with
  summation-by-parts property for the compressible euler equations},
\newblock \bibinfo{journal}{Journal of Computational Physics}
  \bibinfo{volume}{327} (\bibinfo{year}{2016}) \bibinfo{pages}{39--66}.
  \DOIprefix\doi{10.1016/j.jcp.2016.09.013}.
\bibitem[{Flad and Gassner(2017)}]{Flad2017}
\bibinfo{author}{D.~Flad}, \bibinfo{author}{G.~Gassner},
\newblock \bibinfo{title}{On the use of kinetic energy preserving dg-schemes
  for large eddy simulation},
\newblock \bibinfo{journal}{Journal of Computational Physics}
  \bibinfo{volume}{350} (\bibinfo{year}{2017}) \bibinfo{pages}{782--795}.
  \DOIprefix\doi{10.1016/j.jcp.2017.09.004}.
  \href{http://arxiv.org/abs/1706.07601}{{\tt arXiv:1706.07601}}.
\bibitem[{Beck et~al.(2014)Beck, Bolemann, Flad, Frank, Gassner, Hindenlang,
  and Munz}]{Beck2014}
\bibinfo{author}{A.~D. Beck}, \bibinfo{author}{T.~Bolemann},
  \bibinfo{author}{D.~Flad}, \bibinfo{author}{H.~Frank}, \bibinfo{author}{G.~J.
  Gassner}, \bibinfo{author}{F.~Hindenlang}, \bibinfo{author}{C.-D. Munz},
\newblock \bibinfo{title}{High-order discontinuous galerkin spectral element
  methods for transitional and turbulent flow simulations},
\newblock \bibinfo{journal}{International Journal for Numerical Methods in
  Fluids} \bibinfo{volume}{76} (\bibinfo{year}{2014})
  \bibinfo{pages}{522--548}. \DOIprefix\doi{10.1002/fld.3943}.
\bibitem[{Krais et~al.(2021)Krais, Beck, Bolemann, Frank, Flad, Gassner,
  Hindenlang, Hoffmann, Kuhn, Sonntag, and Munz}]{Krais2019}
\bibinfo{author}{N.~Krais}, \bibinfo{author}{A.~Beck},
  \bibinfo{author}{T.~Bolemann}, \bibinfo{author}{H.~Frank},
  \bibinfo{author}{D.~Flad}, \bibinfo{author}{G.~Gassner},
  \bibinfo{author}{F.~Hindenlang}, \bibinfo{author}{M.~Hoffmann},
  \bibinfo{author}{T.~Kuhn}, \bibinfo{author}{M.~Sonntag},
  \bibinfo{author}{C.-D. Munz},
\newblock \bibinfo{title}{Flexi: A high order discontinuous galerkin framework
  for hyperbolic-parabolic conservation laws},
\newblock \bibinfo{journal}{Computers {\&} Mathematics with Applications}
  \bibinfo{volume}{81} (\bibinfo{year}{2021}) \bibinfo{pages}{186--219}.
  \DOIprefix\doi{10.1016/j.camwa.2020.05.004}.
  \href{http://arxiv.org/abs/1910.02858}{{\tt arXiv:1910.02858}}.
\bibitem[{Carpenter and Kennedy(1994)}]{Carpenter1994}
\bibinfo{author}{M.~H. Carpenter}, \bibinfo{author}{a.~Kennedy},
\newblock \bibinfo{title}{Fourth-order kutta schemes},
\newblock \bibinfo{journal}{Nasa Technical Memorandum} \bibinfo{volume}{109112}
  (\bibinfo{year}{1994}) \bibinfo{pages}{1--26}.
\bibitem[{Allievi and Bermejo(1997)}]{Allievi1997}
\bibinfo{author}{A.~Allievi}, \bibinfo{author}{R.~Bermejo},
\newblock \bibinfo{title}{A generalized particle search-locate algorithm for
  arbitrary grids},
\newblock \bibinfo{journal}{Journal of Computational Physics}
  \bibinfo{volume}{132} (\bibinfo{year}{1997}) \bibinfo{pages}{157--166}.
  \DOIprefix\doi{10.1006/jcph.1996.5604}.
\bibitem[{Ortwein et~al.(2019)Ortwein, Copplestone, Munz, Binder, Reschke, and
  Fasoulas}]{Ortwein2019}
\bibinfo{author}{P.~Ortwein}, \bibinfo{author}{S.~M. Copplestone},
  \bibinfo{author}{C.-D. Munz}, \bibinfo{author}{T.~Binder},
  \bibinfo{author}{W.~Reschke}, \bibinfo{author}{S.~Fasoulas},
\newblock \bibinfo{title}{A particle localization algorithm on unstructured
  curvilinear polynomial meshes},
\newblock \bibinfo{journal}{Comput. Phys. Commun.} \bibinfo{volume}{235}
  (\bibinfo{year}{2019}) \bibinfo{pages}{63--74}.
\bibitem[{Wang et~al.(2002)Wang, Shih, and Chang}]{Wang2002}
\bibinfo{author}{S.-W. Wang}, \bibinfo{author}{Z.-C. Shih},
  \bibinfo{author}{R.-C. Chang},
\newblock \bibinfo{title}{An efficient and stable ray tracing algorithm for
  parametric surfaces.},
\newblock \bibinfo{journal}{J. Inf. Sci. Eng.} \bibinfo{volume}{18}
  (\bibinfo{year}{2002}) \bibinfo{pages}{541--561}.
\bibitem[{Yen et~al.(1991)Yen, Spach, Smith, and Pulleyblank}]{Yen1991}
\bibinfo{author}{J.~Yen}, \bibinfo{author}{S.~Spach}, \bibinfo{author}{M.~T.
  Smith}, \bibinfo{author}{R.~W. Pulleyblank},
\newblock \bibinfo{title}{Parallel boxing in b-spline intersection},
\newblock \bibinfo{journal}{{IEEE} Computer Graphics and Applications}
  \bibinfo{volume}{11} (\bibinfo{year}{1991}) \bibinfo{pages}{72--79}.
  \DOIprefix\doi{10.1109/38.67703}.
\bibitem[{Ramsey et~al.(2004)Ramsey, Potter, and Hansen}]{Ramsey2004}
\bibinfo{author}{S.~D. Ramsey}, \bibinfo{author}{K.~Potter},
  \bibinfo{author}{C.~Hansen},
\newblock \bibinfo{title}{Ray bilinear patch intersections},
\newblock \bibinfo{journal}{Journal of Graphics Tools} \bibinfo{volume}{9}
  (\bibinfo{year}{2004}) \bibinfo{pages}{41--47}.
  \DOIprefix\doi{10.1080/10867651.2004.10504896}.
\bibitem[{Nishita et~al.(1990)Nishita, Sederberg, and Kakimoto}]{Nishita1990}
\bibinfo{author}{T.~Nishita}, \bibinfo{author}{T.~W. Sederberg},
  \bibinfo{author}{M.~Kakimoto},
\newblock \bibinfo{title}{Ray tracing trimmed rational surface patches},
\newblock in: \bibinfo{booktitle}{Proceedings of the 17th annual conference on
  Computer graphics and interactive techniques - SIGGRAPH '90}, SIGGRAPH '90,
  \bibinfo{publisher}{ACM Press}, \bibinfo{address}{New York, New York, USA},
  \bibinfo{year}{1990}, pp. \bibinfo{pages}{337--345}.
  \DOIprefix\doi{10.1145/97879.97916}.
\bibitem[{Fasoulas et~al.(2019)Fasoulas, Munz, Pfeiffer, Beyer, Binder,
  Copplestone, Mirza, Nizenkov, Ortwein, and Reschke}]{Fasoulas2019}
\bibinfo{author}{S.~Fasoulas}, \bibinfo{author}{C.-D. Munz},
  \bibinfo{author}{M.~Pfeiffer}, \bibinfo{author}{J.~Beyer},
  \bibinfo{author}{T.~Binder}, \bibinfo{author}{S.~Copplestone},
  \bibinfo{author}{A.~Mirza}, \bibinfo{author}{P.~Nizenkov},
  \bibinfo{author}{P.~Ortwein}, \bibinfo{author}{W.~Reschke},
\newblock \bibinfo{title}{Combining particle-in-cell and direct simulation
  monte carlo for the simulation of reactive plasma flows},
\newblock \bibinfo{journal}{Physics of Fluids} \bibinfo{volume}{31}
  (\bibinfo{year}{2019}) \bibinfo{pages}{072006}.
  \DOIprefix\doi{10.1063/1.5097638}.
\bibitem[{Pfeiffer(2018)}]{Pfeiffer2018a}
\bibinfo{author}{M.~Pfeiffer},
\newblock \bibinfo{title}{Particle-based fluid dynamics: Comparison of
  different bhatnagar-gross-krook models and the direct simulation monte carlo
  method for hypersonic flows},
\newblock \bibinfo{journal}{Physics of Fluids} \bibinfo{volume}{30}
  (\bibinfo{year}{2018}) \bibinfo{pages}{106106}.
  \DOIprefix\doi{10.1063/1.5042016}.
\bibitem[{Kopper et~al.(2022)Kopper, Copplestone, Pfeiffer, Koch, Fasoulas, and
  Beck}]{Kopper2022}
\bibinfo{author}{P.~Kopper}, \bibinfo{author}{S.~M. Copplestone},
  \bibinfo{author}{M.~Pfeiffer}, \bibinfo{author}{C.~Koch},
  \bibinfo{author}{S.~Fasoulas}, \bibinfo{author}{A.~Beck},
\newblock \bibinfo{title}{Hybrid parallelization of
  {E}uler{\textendash}{L}agrange simulations based on {MPI}-3 shared memory},
\newblock \bibinfo{journal}{Advances in Engineering Software}
  \bibinfo{volume}{174} (\bibinfo{year}{2022}) \bibinfo{pages}{103291}.
  \DOIprefix\doi{10.1016/j.advengsoft.2022.103291}.
\bibitem[{Group(2022)}]{hdf5}
\bibinfo{author}{T.~H. Group}, \bibinfo{title}{Hierarchical data format,
  version 5}, \bibinfo{year}{1997-2022}.
  \bibinfo{note}{Https://www.hdfgroup.org/HDF5/}.
\bibitem[{Forum(2021)}]{mpi40}
\bibinfo{author}{M.~P.~I. Forum}, \bibinfo{title}{{MPI}: A Message-Passing
  Interface Standard Version 4.0}, \bibinfo{year}{2021}. \URLprefix
  \url{https://www.mpi-forum.org/docs/mpi-4.0/mpi40-report.pdf}.
\bibitem[{Ortwein et~al.(2018)Ortwein, Binder, Copplestone, Mirza, Nizenkov,
  Pfeiffer, Munz, and Fasoulas}]{Ortwein2018}
\bibinfo{author}{P.~Ortwein}, \bibinfo{author}{T.~Binder},
  \bibinfo{author}{S.~Copplestone}, \bibinfo{author}{A.~Mirza},
  \bibinfo{author}{P.~Nizenkov}, \bibinfo{author}{M.~Pfeiffer},
  \bibinfo{author}{C.-D. Munz}, \bibinfo{author}{S.~Fasoulas},
\newblock \bibinfo{title}{A load balance strategy for hybrid particle-mesh
  methods},
\newblock \bibinfo{journal}{arXiv preprint arXiv:1811.05152}
  (\bibinfo{year}{2018}).
\bibitem[{Schamberger and Wierum(2003)}]{Schamberger2003}
\bibinfo{author}{S.~Schamberger}, \bibinfo{author}{J.-M. Wierum},
\newblock \bibinfo{title}{Graph partitioning in scientific simulations:
  Multilevel schemes versus space-filling curves},
\newblock in: \bibinfo{booktitle}{Lecture Notes in Computer Science},
  \bibinfo{publisher}{Springer Berlin Heidelberg}, \bibinfo{year}{2003}, pp.
  \bibinfo{pages}{165--179}. \DOIprefix\doi{10.1007/978-3-540-45145-7\_14}.
\bibitem[{Mitchell(2007)}]{Mitchell2007}
\bibinfo{author}{W.~F. Mitchell},
\newblock \bibinfo{title}{A refinement-tree based partitioning method for
  dynamic load balancing with adaptively refined grids},
\newblock \bibinfo{journal}{Journal of Parallel and Distributed Computing}
  \bibinfo{volume}{67} (\bibinfo{year}{2007}) \bibinfo{pages}{417--429}.
  \DOIprefix\doi{10.1016/j.jpdc.2006.11.003}.
\bibitem[{Ahrens et~al.(2005)Ahrens, Geveci, and Law}]{ahrens2005paraview}
\bibinfo{author}{J.~Ahrens}, \bibinfo{author}{B.~Geveci},
  \bibinfo{author}{C.~Law},
\newblock \bibinfo{title}{Paraview: An end-user tool for large data
  visualization},
\newblock \bibinfo{journal}{The visualization handbook} \bibinfo{volume}{717}
  (\bibinfo{year}{2005}).
\bibitem[{Beck et~al.(2016)Beck, Flad, Tonh{\"{a}}user, Gassner, and
  Munz}]{Beck2016}
\bibinfo{author}{A.~D. Beck}, \bibinfo{author}{D.~G. Flad},
  \bibinfo{author}{C.~Tonh{\"{a}}user}, \bibinfo{author}{G.~Gassner},
  \bibinfo{author}{C.-D. Munz},
\newblock \bibinfo{title}{On the influence of polynomial de-aliasing on subgrid
  scale models},
\newblock \bibinfo{journal}{Flow, Turbulence and Combustion}
  \bibinfo{volume}{97} (\bibinfo{year}{2016}) \bibinfo{pages}{475--511}.
  \DOIprefix\doi{10.1007/s10494-016-9704-y}.
\bibitem[{Armenio and Fiorotto(2001)}]{Armenio2001}
\bibinfo{author}{V.~Armenio}, \bibinfo{author}{V.~Fiorotto},
\newblock \bibinfo{title}{The importance of the forces acting on particles in
  turbulent flows},
\newblock \bibinfo{journal}{Physics of Fluids} \bibinfo{volume}{13}
  (\bibinfo{year}{2001}) \bibinfo{pages}{2437--2440}. \URLprefix
  \url{http://aip.scitation.org/doi/10.1063/1.1385390}.
  \DOIprefix\doi{10.1063/1.1385390}.
\bibitem[{Haugen and Kragset(2010)}]{Haugen2010}
\bibinfo{author}{N.~E.~L. Haugen}, \bibinfo{author}{S.~Kragset},
\newblock \bibinfo{title}{Particle impaction on a cylinder in a crossflow as
  function of {S}tokes and {R}eynolds numbers},
\newblock \bibinfo{journal}{Journal of Fluid Mechanics} \bibinfo{volume}{661}
  (\bibinfo{year}{2010}) \bibinfo{pages}{239--261}.
  \DOIprefix\doi{10.1017/s0022112010002946}.
\bibitem[{Muhr(1976)}]{Muhr1976}
\bibinfo{author}{W.~Muhr}, \bibinfo{title}{Theoretical and experimental
  investigation of particle deposition in fibrous filters by field and inertial
  forces}, Ph.D. thesis, Institut f{\"{u}}r Mechanische Verfahrenstechnik und
  Mechanik, Universit{\"{a}}t Karlsruhe, \bibinfo{address}{Karlsruhe, Germany},
  \bibinfo{year}{1976}.
\bibitem[{Kuhn et~al.(2020)Kuhn, Kempf, Beck, and Munz}]{Kuhn2020}
\bibinfo{author}{T.~Kuhn}, \bibinfo{author}{D.~Kempf},
  \bibinfo{author}{A.~Beck}, \bibinfo{author}{C.-D. Munz},
\newblock \bibinfo{title}{A novel turbulent inflow method for zonal large eddy
  simulations with a discontinuous {G}alerkin solver},
\newblock \bibinfo{journal}{Submitted to Computers and Fluids}
  (\bibinfo{year}{2020}).
\bibitem[{Dunn et~al.(1996)Dunn, Baran, and Miatech}]{Dunn1996}
\bibinfo{author}{M.~G. Dunn}, \bibinfo{author}{A.~J. Baran},
  \bibinfo{author}{J.~Miatech},
\newblock \bibinfo{title}{Operation of gas turbine engines in volcanic ash
  clouds},
\newblock \bibinfo{journal}{Journal of Engineering for Gas Turbines and Power}
  \bibinfo{volume}{118} (\bibinfo{year}{1996}) \bibinfo{pages}{724--731}.
  \DOIprefix\doi{10.1115/1.2816987}.
\bibitem[{Hillewaert et~al.(2013)Hillewaert, de~Wiart, and
  Arts}]{Hillewaert2013}
\bibinfo{author}{K.~Hillewaert}, \bibinfo{author}{C.~C. de~Wiart},
  \bibinfo{author}{T.~Arts},
\newblock \bibinfo{title}{Dns and les of transitional flow around a high lift
  turbine cascade at low reynolds number},
\newblock in: \bibinfo{booktitle}{2nd International Workshop on High-Order CFD
  Methods}, \bibinfo{organization}{von Karman Institute, Turbomachinery
  Department}, \bibinfo{publisher}{DLR, AIAA and AFOSR},
  \bibinfo{address}{Cologne, Germany}, \bibinfo{year}{2013}.
\bibitem[{Kawamura et~al.(1984)Kawamura, Hiwada, Hibino, Mabuchi, and
  Kumada}]{Kawamura1984}
\bibinfo{author}{T.~Kawamura}, \bibinfo{author}{M.~Hiwada},
  \bibinfo{author}{T.~Hibino}, \bibinfo{author}{I.~Mabuchi},
  \bibinfo{author}{M.~Kumada},
\newblock \bibinfo{title}{Flow around a finite circular cylinder on a flat
  plate : Cylinder height greater than turbulent boundary layer thickness},
\newblock \bibinfo{journal}{Bulletin of {JSME}} \bibinfo{volume}{27}
  (\bibinfo{year}{1984}) \bibinfo{pages}{2142--2151}.
  \DOIprefix\doi{10.1299/jsme1958.27.2142}.

\end{thebibliography}

\end{document}